\documentclass[aps,prd,twocolumn,superscriptaddress]{revtex4-2}
\usepackage{bm}
\usepackage{slashed}
\usepackage{amsmath}
\usepackage[T1]{fontenc}
\usepackage{hyperref}
\usepackage{graphicx}
\usepackage{amssymb}
\usepackage{comment}

\newcommand{\kP}{\kappa_{\mathrm{P}}}
\newcommand{\kIR}{\kappa_{\mathrm{IR}}}
\newcommand{\ktIR}{\tilde k_{\mathrm{IR}}}
\newcommand{\KE}{K_{\mathrm{E}}}
\newcommand{\lP}{l_{\mathrm{P}}}
\newcommand{\RH}{R_{\mathrm{H}}}
\newcommand{\kc}{\kappa_{\mathrm{c}}}
\newcommand{\kend}{\kappa_{\mathrm{end}}}

\begin{document}
\title{Self-Gravitation of Mode Quanta in a Causal Resonator: One-Loop Finiteness in Linearized Quantum Gravity and the Emergence of the Dark-Energy Scale}
\author{Christian Rembe}
\email{christian.rembe@tu-clausthal.de}
\affiliation{Clausthal Technical University, Institute of Electrical Information Technology, Leibnizstr. 28\\38678 Clausthal-Zellerfeld, Germany}
\date{September 2026}
\begin{abstract}
Perturbative quantum gravity is ultraviolet divergent and, as shown by 't Hooft and Veltman, non-renormalizable. A recent object-relative, Lorentz-invariant weighting of internal electromagnetic modes renders selected one-loop contributions of quantum electrodynamics finite without counterterms. Here that weighting is derived rather than postulated: causality defines a mode resonator bounded by the Hubble radius and re-defined in every inertial frame, and the self-gravitation of each mode quantum deforms its spectrum. Applied to linearized gravity, the graviton self-coupling reaches order unity at the Planck wave number, so that the mode content available to a loop is suppressed beyond it. Any suppression exceeding a threshold fixed by the diagram renders the one-loop integrals finite; no cutoff is imposed and no mode structure above the Planck scale need be derived. For pure gravity the same power counting extends to every loop order, and the leading suppression already meets the resulting condition for every topology with more than one vertex; a single class of single-vertex insertions is excluded. The Bianchi identity takes the role of the Ward identity at one loop, and the transformation between the resonators of two inertial frames is unitary. The same finite mode content yields a vacuum energy density of order $H^{2}/G$, of the observed sign and magnitude, from the Planck and causal infrared wave numbers alone. A contribution tracking the instantaneous expansion rate is an exact rescaling of the gravitational constant and is excluded by primordial nucleosynthesis, so the mode basis must be frozen in any inertial system; referring it to the stationary de Sitter horizon fixes the remaining coefficient to the geometric value $6/\pi$ and makes a value for the energy density of gravitational vacuum fluctuations possible which can be interpreted as origin of the dark-energy effect.
\end{abstract}
\maketitle
\section{Introduction}
\label{sec:intro}
The quantization of gravity remains one of the central unsolved problems of theoretical physics. A first systematic attempt at a canonical formulation was given by DeWitt \cite{DeWitt1967}, leading to the Wheeler-DeWitt equation $\mathbf{H}|\Psi\rangle = 0$. While this approach captures the formal structure of quantum geometrodynamics, it introduces fundamental difficulties with ultraviolet regularization that have not been resolved within the canonical framework.
A complementary perturbative approach treats the gravitational field as a linearized spin-2 perturbation $h_{\mu\nu}$ around flat spacetime and quantizes the resulting field theory. In this setting, 't Hooft and Veltman showed that one-loop divergences arise in the presence of matter fields that cannot be absorbed by renormalization of the original parameters of the Einstein-Hilbert action \cite{tHooft1974}. Unlike renormalizable gauge theories, where a finite number of counterterms suffices to absorb all ultraviolet divergences at every loop order, the gravitational coupling constant $G$ carries negative mass dimension in natural units. This causes the degree of divergence to increase with each loop order, requiring an infinite number of independent counterterms and rendering the theory perturbatively non-renormalizable. At two-loop level, Goroff and Sagnotti demonstrated that this divergence structure persists even in pure gravity without matter, confirming the fundamental nature of the problem \cite{Goroff1986}.
Within the effective field theory framework developed by Donoghue \cite{Donoghue1994}, linearized quantum gravity can be treated as a consistent low-energy theory. Physical observables such as quantum corrections to the Newtonian potential are well-defined and finite at long distances, since the infrared structure of the theory is unaffected by the ultraviolet problem. However, the ultraviolet divergences themselves remain, and the theory retains its non-renormalizable character. Alternative approaches such as asymptotic safety \cite{Reuter1998} seek a non-trivial ultraviolet fixed point of the gravitational couplings, but a complete and broadly accepted resolution has not been established. Within Wilson's renormalization group framework, ultraviolet divergences are handled by integrating out high-frequency modes and absorbing their effects into running coupling constants, with the cutoff scale treated as an external parameter~\cite{Wilson1971a, Wilson1971b, Wilson1974, Weinberg1979}.
The present work is based on the heuristic possibility that sufficiently localized field modes may not remain spectrally inert with respect to the localization conditions defining the interaction itself. The central hypotheses of this work are that the continuum density of plane-wave modes is not the fundamental physical mode spectrum associated with an inertial observer in an infinitely large space and that the gravitation of the energy of a mode quantum is not negligible. Thus,
\begin{enumerate}
 \item physical mode spectra are constrained by causal coherence, and
 \item for gravitational modes the energy carried by the modes further modifies their spectrum through self-gravitational backreaction.
\end{enumerate}
A third assumption enters and does not follow from these two: that the causal boundary is constructed identically in every inertial frame, so that the same spectrum is obtained in each. Since that boundary is fixed by $\RH=c/H_{0}$ and $H_{0}$ is defined through the isotropic expansion, this concerns the frame-independence of the construction rather than the constancy of $c$, and it is stated as a postulate in Sec.~\ref{subsubsec:invariant_postulate}.
 In the accompanying object-relative mode framework, this is modeled by an effective covariant mode weighting depending on the mode frequency measured in the rest frame of the localized interaction object, $u^\mu k_\mu$, leading asymptotically to a thinning of the effective mode density towards higher frequencies. In the gravitational case, such an interpretation becomes particularly natural, since gravity couples universally to energy, including the localization energy and self-interaction of the modes themselves. The resulting modification of the mode structure is therefore not introduced here as an external regulator prescription, but as an effective manifestation of gravitational backreaction within a spherical finite volume which excludes the space beyond the Hubble radius because it is not causally connected to the considered inertial system.

 A recent paper \cite{Rembe2026} introduced such a framework in an infinite space as an object-relative ultraviolet weighting of internal electromagnetic modes in quantum electrodynamics in an infinite space.  There, the weighting suppresses ultraviolet contributions asymptotically as $(c \, k_c)^3/(u \, k)^3$ relative to the standard four-dimensional loop mode density, while leaving the infrared sector unchanged below the characteristic scale expressed by the four vector $k_c = (\omega_c /c, \bm{\kappa}_c)$, which is set by the localization of the interaction. Under this weighting, selected one-loop QED contributions become ultraviolet finite without counterterms, a restricted one-loop Ward consistency check is preserved, and the characteristic scales emerging from the anomalous magnetic moment and a Bethe-type Lamb-shift estimate are found to lie near physically motivated localization scales of the respective interactions. Also the Casimir effect does not change in the new framework.
In the present article, this object-relative mode weighting is applied to the linearized gravitational field and it is extended to a theory involving a finite-mode resonator. The great difference to the previous approach is the much longer coherence length of gravitational modes due to the much smaller coupling constant compared to other fields. Other fields interact so strongly and so locally that approximating a very large finite volume as infinite remains valid. This approximation cannot be made if coherence is much longer than the considered volume.
The graviton propagator in De Donder gauge \cite{Misner1973} is recalled, and the weighting is introduced for internal graviton modes in the same way as for photon modes in \cite{Rembe2026}. For internal graviton lines the characteristic scale is the Planck wave number, since a purely gravitational loop contains no external localization scale. A Euclidean power-counting argument then shows that the one-loop integrals are finite provided the mode content above that scale is suppressed more strongly than a threshold fixed by the diagram; that threshold lies far below the value the construction supplies, so the conclusion rests on the existence of a suppression rather than on its form. The scale thus follows from the construction rather than being imposed externally, and the conclusion does not depend on the detailed mode structure above it. For pure gravity the same counting extends to every loop order, covering every topology with more than one vertex and leaving a single class of single-vertex insertions outside. The consistency of the weighting with the Bianchi identity is examined at one-loop level, in direct analogy to the Ward consistency check of \cite{Rembe2026}. A first characteristic scale, the infrared Hubble wave number $\tilde{k}_{IR}=\frac{1}{4 R_H}=\frac{ H_{0}}{4c}$, with the Hubble constant $H_0$ and the Hubble radius $R_H$ enters naturally through the cosmological interaction volume. A second characteristic ultraviolet crossover scale emerges as the Planck wave number
  $\kP = 1/\lP = \sqrt{c^{3}/\hbar G}$. Together, these two parameter-free scales determine an effective finite vacuum energy density. Throughout this work, a tilde denotes the ordinary (spectroscopic) wavenumber, $\tilde k = 1/\lambda$, and $\kappa$ denotes the corresponding reduced wavenumber, $\kappa = 2\pi/\lambda = \omega/c$, related by $\kappa = 2\pi\tilde k$. The loop integrals of Sec.~\ref{sec:powercounting} and the object-relative variable $u\cdot k$ with four velocity $u$ and four wave vector $k$ are formulated throughout in terms of the reduced wavenumber $\kappa$.
The present approach differs in that the modification of the high-frequency mode structure arises from a physical mechanism — the gravitational self-backreaction of the modes themselves. The present work does not assume the weighting as a purely external regulator prescription, but interprets it as an effective manifestation of gravitational self-interaction at sufficiently localized scales \cite{Rembe2026}.

In the standard treatment, the vacuum energy density of quantum fields diverges as $\rho_{\text{vac}} \sim \int \mathrm{d}^3\kappa\, \hbar c \kappa \to \infty$, exceeding the observed dark energy density by approximately 120 orders of magnitude with a cutoff set at the Planck scale. Within the object-relative mode framework proposed here, the coherent mode content of the causal resonator, which is governed by the infrared scale rather than by the local scale relevant to loop corrections, leads naturally to an effective vacuum energy scale of order $\rho_{vac} \sim \frac{c^2 H_0^2}{G}$. Because the causal resonator is constructed identically in every inertial frame and from Lorentz scalars alone, every inertial observer assigns this contribution the same energy density, which forces its pressure to be negative and equal in magnitude to that density. The term is therefore repulsive: it accelerates the expansion of the universe and yields a value of order $H_0^2/G$, which is in the order of magnitude of the observed dark energy \cite{Peebles2003}.
The findings suggest that gravitational self-backreaction on the effective structure of a  mode resonator defined by causality may capture part of the infrared physics underlying the observed cosmological vacuum scale. Although not constituting a full cosmological model, the emergence of the observed vacuum-density scale from the same ultraviolet mode-structure mechanism provides a nontrivial infrared consistency check of the framework.

The scope and limitations of the approach to quantizing the linearized field equations are discussed and directions for further work are indicated. The effective field theory framework, developed by Wilson~\cite{Wilson1974} and Weinberg~\cite{Weinberg1979} and applied to gravity by Donoghue~\cite{Donoghue1994}, treats the ultraviolet cutoff as an external parameter. The present approach is complementary: the scale at which the graviton mode content terminates follows from the construction itself, so that the loop integrals are bounded without an externally chosen regulator. The calculation of the density of the gravitational vacuum energy is possible within this framework. However, it is not understood yet how vacuum fluctuations dissipate by interaction with other quantum fields. The residual coefficient of order unity separating the estimate from the measured value is not fitted but follows from a fixed-point condition, and it is consistent with the coherent spectrum ending just above the Planck wave number.
\section{The Graviton Propagator in Linearized Gravity}
\label{sec:propagator}
In the linearized theory, the gravitational field is described as a small symmetric tensor perturbation $h_{\mu\nu}$ around flat Minkowski spacetime \cite{Misner1973},
\begin{equation}
  g_{\mu\nu} = \eta_{\mu\nu} + h_{\mu\nu},
  \qquad
  |h_{\mu\nu}| \ll 1.
  \label{eq:linearization}
\end{equation}
This decomposition defines the weak-field regime in which the gravitational field is treated as a quantum field on a fixed flat background. The dynamics of $h_{\mu\nu}$ are governed by the Einstein--Hilbert action expanded to second order in $h_{\mu\nu}$. To fix the gauge freedom associated with linearized diffeomorphisms, the De~Donder gauge condition
\begin{equation}
  \partial^{\nu} h_{\mu\nu} = \frac{1}{2}\,\partial_{\mu} h,
  \qquad
  h = \eta^{\mu\nu} h_{\mu\nu},
  \label{eq:dedonder}
\end{equation}
is imposed. Substituting Eq.~\eqref{eq:linearization} into the Einstein field equations and expanding to first order in $h_{\mu\nu}$, one obtains in De~Donder gauge~\eqref{eq:dedonder} the linearized field equation
\begin{equation}
  \Box\,\bar{h}_{\mu\nu} = -\frac{16\pi G}{c^{4}}\,T_{\mu\nu},
  \label{eq:linearized_field}
\end{equation}
where $\bar{h}_{\mu\nu} = h_{\mu\nu} - \tfrac{1}{2}\eta_{\mu\nu}h$ is the trace-reversed perturbation and $\Box = \eta^{\mu\nu}\partial_{\mu}\partial_{\nu}$ is the d'Alembert operator. Upon canonical quantization, $\bar{h}_{\mu\nu}$ and $T_{\mu\nu}$ are replaced by field operators:
\begin{equation}
  \Box\,\hat{\bar{h}}_{\mu\nu} = -\frac{16\pi G}{c^{4}}\,\hat{T}_{\mu\nu}.
  \label{eq:quantized_field}
\end{equation}
The left-hand side of Eq.~\eqref{eq:quantized_field} describes the quantum fluctuations of the gravitational field itself — the graviton field operator $\hat{\bar{h}}_{\mu\nu}$. The right-hand side contains the energy-momentum tensor operator $\hat{T}_{\mu\nu}$ of all matter and gauge fields of the Standard Model. What the vacuum fluctuations of the former contribute to the expansion is not decided by this placement, which fixes only a relative sign, but by their tensor structure, which is determined in Sec.~\ref{subsec:geometric}.
In De~Donder gauge, the graviton propagator takes the form~\cite{tHooft1974}
\begin{equation}
  D_{\mu\nu\rho\sigma}(k) = \frac{-\,i\, P_{\mu\nu\rho\sigma}}{k^{2}},
  \label{eq:graviton_propagator}
\end{equation}
where the spin-2 numerator tensor is
\begin{equation}
  P_{\mu\nu\rho\sigma} = \frac{1}{2}\!\left(
      \eta_{\mu\rho}\eta_{\nu\sigma}
    + \eta_{\mu\sigma}\eta_{\nu\rho}
    - \eta_{\mu\nu}\eta_{\rho\sigma}
    \right).
  \label{eq:spin2_tensor}
\end{equation}
This is the spin-2 analogue of the photon propagator in Feynman gauge. The scalar ultraviolet behavior of the graviton propagator is governed by the same factor $1/k^{2}$ with the four wave vector $k$ as the photon propagator. The spin-2 structure encoded in $P_{\mu\nu\rho\sigma}$ is algebraic and does not affect the ultraviolet power counting of the loop integrals.
The graviton couples to the full energy-momentum tensor $T_{\mu\nu}$ of all fields present, including its own gravitational contribution. The interaction vertex is proportional to the gravitational coupling $\kappa_{\mathrm g} = \sqrt{16\pi G/c^{4}}$ and introduces numerator factors that grow with the loop momentum $k$. For a one-loop diagram with two graviton propagators and one matter propagator, the Euclidean integrand behaves asymptotically as
\begin{equation}
  F(\KE) \sim \frac{\KE^{2}}{\KE^{4}} = \frac{1}{\KE^{2}},
  \label{eq:integrand_asymptotic}
\end{equation}
where the factor $\KE^{2}$ in the numerator arises from the gravitational vertex structure, and the factor $\KE^{4}$ in the denominator comes from the two graviton propagators. In four Euclidean dimensions, the radial integration measure contributes an additional factor $\KE^{3}$, so that the unweighted one-loop integral behaves as
\begin{equation}
  I \sim \int \mathrm{d}\KE\; \KE^{3}\, F(\KE)
    \sim \int \mathrm{d}\KE\; \KE
    \;\longrightarrow\; \infty.
  \label{eq:uv_divergence}
\end{equation}
This linear ultraviolet divergence cannot be absorbed by renormalization of the original parameters of the Einstein--Hilbert action. As shown by 't\,Hooft and Veltman~\cite{tHooft1974}, the divergence structure requires a counterterm of a form not present in the original action, and at each successive loop order new independent counterterms of increasing order in the curvature are required. The theory is therefore perturbatively non-renormalizable. At two-loop order, Goroff and Sagnotti confirmed that this divergence structure persists in pure gravity without matter~\cite{Goroff1986}, establishing the fundamental nature of the problem.
Within the effective field theory framework of Donoghue~\cite{Donoghue1994}, the ultraviolet problem does not prevent the calculation of finite quantum corrections to long-range gravitational observables. The one-loop calculation yields a finite quantum correction to the two-body Newtonian potential \emph{energy} of two masses $m$ and $M$~\cite{Donoghue1994,BjerrumBohr2003},
\begin{equation}
  V(r) = -\,\frac{G m M}{r}
  \left(
    1 + 3\,\frac{G(m+M)}{r c^{2}} + \frac{41}{10\pi}\,\frac{G\hbar}{r^{2} c^{3}} + \cdots
  \right),
  \label{eq:newtonian_corrections}
\end{equation}
where the classical post-Newtonian coefficient $3(m+M)$ and the quantum coefficient $41/(10\pi)$ are fixed, calculable numbers~\cite{BjerrumBohr2003} (superseding an earlier, later-corrected coefficient in Ref.~\cite{Donoghue1994}), independent of the ultraviolet divergences of Sec.~\ref{sec:propagator}. The ultraviolet problem therefore does not affect these infrared predictions of the theory, but leaves the short-distance behavior fundamentally unresolved.
Since the $\hbar$-dependent term in Eq.~\eqref{eq:newtonian_corrections} carries no dependence on $m$ or $M$ beyond the overall prefactor $Gm M/r$, dividing by one mass ($m$ for a probe mass) results in the ordinary one-body gravitational \emph{potential},
\begin{equation}
  \Phi(r) = -\,\frac{G M}{r}
  \left(
    1 + 3\,\frac{G M}{r c^{2}} + \frac{41}{10\pi}\,\frac{G\hbar}{r^{2} c^{3}} + \cdots
  \right).
  \label{eq:donoghue_potential}
\end{equation}
The potential is used in the next sections to derive a mode structure.
The next chapter introduces the two hypotheses which extend the fundamental assumptions of quantum field theories. The results of this article follow from them together with the invariance postulate of Sec.~\ref{subsubsec:invariant_postulate}, in a chain of conclusions whose remaining undetermined elements --- the normalization $\mathcal{N}_{H}$, the object-relative assignment of $\kc$, and the behaviour of the weighting at the nonlinear vertices --- are identified where they enter.

\section{A New Relativity Principle for Quantum Fields and Self-Gravitation of the Mode Quanta}
\label{sec:weighting}
\subsection{The causal resonator with self-gravitation of the mode quanta}
\label{subsec:causal_resonator}

\subsubsection{Hypothesis 1: the Hubble radius as a causal boundary}

In the previous article~\cite{Rembe2026}, the mode spectrum of the inertial system was discussed rather heuristically and concluded from the gravitational self-energy of a quantum of any mode and an overlap argument. This seems to be a valid assumption for material fields and gravitons interacting locally. However, the coupling of gravity is extremely weak and it can be assumed that not all gravitons of the vacuum fluctuations are absorbed by interactions. Thus, the question arises of whether the mode spectrum of the vacuum exists as a well-defined physical quantity for any inertial system. Space is permanently expanding, driven in the present framework by the repulsive gravitational vacuum energy identified below. For any inertial observer, this expansion defines a causal boundary: objects beyond the Hubble radius
\begin{equation}
  \RH = \frac{c}{H_{0}}
  \label{eq:hubble_radius}
\end{equation}
recede faster than the speed of light and cannot be causally connected to the observer. The Hubble radius approximates the event horizon in an accelerating universe: beyond this boundary, signals emitted today will never reach the observer. No mode of any field — including the gravitational field itself — can exist in a given inertial frame without respecting this boundary.

Two properties of the admissible mode spectrum follow, and both are obtained without imposing any condition on the field amplitude at the horizon. The horizon is a causal limit, not a reflecting wall. A node condition would in fact be unphysical here: excitations do cross the Hubble horizon as the universe expands, and enforcing their reflection would prevent precisely the outward transit that the expanding causal structure requires. The mode phases therefore remain unconstrained throughout.

The first property is the length scale of the relevant causal structure. Two points belong to one coherent mode only if a phase relation between them can be established and maintained, and this requires a \emph{symmetric} causal relation: each point must lie in the past light cone of the other. One-way signalling is insufficient, since it allows one point to be informed about the other but does not define a phase relation agreed between them. Mutual causal contact across a separation $\ell$ therefore corresponds to a closed causal path of length $2\ell$. The largest separation contained in the causal sphere is its diameter $D_H = 2\RH$, so that the closed causal path characterizing the region has the length $2D_H = 4\RH$.

The second property is the discreteness of the spectrum, and it follows from a uniqueness requirement on the basis. In every inertial frame the basis states of a field must permit, for any excitation of that field, an unambiguous assignment of occupation numbers: a quantum may be in a superposition of basis states, but the basis in which that superposition is expressed must itself be unique. A wave whose wavelength is not commensurate with the closed causal path is not an independent degree of freedom in this sense, since on that path it is already representable as a superposition of the commensurate ones; admitting it as a separate basis element would allow the same physical excitation to possess more than one decomposition. Uniqueness of the basis therefore restricts the admissible wavenumbers to those commensurate with the closed causal path, which is a periodicity condition on that path rather than a boundary condition on the amplitude. The admissible wavelengths are consequently
\begin{equation}
  \lambda_{n} = \frac{4\RH}{n+1}, \qquad n = 0, 1, 2, 3, \ldots,
  \;\; \lambda_{0}  = 4\RH,
  \label{eq:lambda_0}
\end{equation}
with $\lambda_{0}$ the longest admissible wavelength.
The characteristic infrared scale of the causal resonator is defined from this fundamental wavelength~\eqref{eq:lambda_0}, rather than from the Hubble diameter directly, since $\lambda_0$ is the longest wavelength permitted by the causal-contact requirement:
\begin{equation}
  \ktIR \equiv \frac{1}{\lambda_{0}} = \frac{1}{4\RH} = \frac{H_{0}}{4c},
  \; \;
  \kIR = 2\pi\,\ktIR = \frac{\pi H_{0}}{2c}.
  \label{eq:kIR_def}
\end{equation}
\subsubsection{Hypothesis 2: self-gravitation in the resonator of a fixed inertial frame}

A mode of wavenumber $\kappa$ carries energy $E = \hbar c\kappa$ and an associated gravitating mass $m = \hbar\kappa/c$. A coherent eigenmode of the resonator is delocalized over its entire volume irrespective of its wavelength: the wavelength sets the scale on which the mode oscillates, not its spatial extent, and spatially localized states arise only from superpositions of many modes, which requires the interaction-induced decoherence discussed in Sec.~\ref{subsec:loss_origin} — such decohered modes being precisely the ones excluded from the coherent vacuum contribution evaluated in Sec.~\ref{sec:darkenergy}. The mass $M=m$ associated with a mode may therefore be treated as distributed over the whole resonator, in the same sense in which the standard vacuum energy $\rho_{\mathrm{vac}}=V^{-1}\sum_{k}\hbar\omega_{k}/2$ assigns the energy of each mode to the entire quantization volume; for $\kappa\gg\kIR$ the oscillating mode profile averages to a uniform density over any region large compared with $1/\kappa$. Placed in the fixed causal resonator, this mass contributes to the gravitational potential at the center of the sphere which coincides with the inertial system,
\begin{equation}
  \delta\Phi = -\frac{3Gm}{2\RH}
  = -\frac{3G\hbar\kappa}{2c\RH}
  = -\frac{3GH_{0}\hbar\kappa}{2c^{2}},
  \label{eq:phi_center}
\end{equation}
using the standard result for the Newtonian potential of an inertial system at the center of a uniform sphere of radius $\RH$. This changes the total energy of the quantum by its own gravitational self-energy,
\begin{equation}
  \delta E(\kappa) = m\,\delta\Phi
  = -\frac{3GH_{0}\hbar^{2}\kappa^{2}}{2c^{3}},
  \label{eq:delta_E}
\end{equation}
and correspondingly shifts its wavenumber by
\begin{equation}
\begin{aligned}
  \delta \kappa(\kappa)
  &= \frac{\delta E(\kappa)}{\hbar c}
  = -\frac{3GH_{0}\hbar\kappa^{2}}{2c^{4}}
  = -\alpha\,\kappa^{2},
  \\
  \alpha &= \frac{3GH_{0}\hbar}{2c^{4}} = \frac{3}{\pi}\cdot\frac{\kIR}{\kP^{2}}.
\end{aligned}
  \label{eq:delta_k_IR}
\end{equation}
Equation~\eqref{eq:delta_E} evaluates the energy of the mode in the
potential~\eqref{eq:phi_center} at the centre of the resonator, which
is the observation point of the inertial frame defining it. This is
the operationally relevant quantity: what the recursion tracks is the
spectrum as reckoned at that point, and since every inertial frame
constructs its resonator about its own centre, the prescription is the
same in all of them. The coefficient of the chain
density~\eqref{eq:dn_dk} is in any case not propagated into the
normalization $\mathcal{N}_H$ of Sec.~\ref{subsec:normalization} --
that normalization is carried by the coefficient $K_{gr}$ introduced
there -- and the $\kappa^{-2}$ dependence, which is the robust content
of the construction, is independent of it.

Note that $\alpha\kP \sim \kIR/\kP \sim 10^{-61}$, so the correction is parametrically small at every scale below $\kP$. This is the same physical mechanism — energy gravitates, and its self-energy shifts the mode's wavenumber — that underlies the mode weighting of Ref.~\cite{Rembe2026}; the only difference is the resonator scale entering the potential: here it is the fixed Hubble radius $\RH$, rather than the mode's own wavelength.

Equation~\eqref{eq:phi_center} uses only the leading-order Newtonian potential. Using instead the effective-field-theory-corrected potential of a point source, Eq.~\eqref{eq:donoghue_potential}, with source mass $M=m=\hbar\kappa/c$ and $r=\RH$, the classical post-Newtonian correction relative to the leading term is of order
\begin{equation}
  \frac{Gm(\kappa)}{\RH c^{2}} \sim \frac{2}{\pi}\,\frac{\kappa\,\kIR}{\kP^{2}},
  \label{eq:PN_ratio}
\end{equation}
which reaches at most $\sim\kIR/\kP\sim10^{-61}$ even at $\kappa=\kP$ (where $m(\kappa)$ equals the Planck mass): the correction stays small not because the source mass is small, but because it is compared to the vastly larger distance $\RH$ rather than to its own scale $1/\kappa$. The genuinely quantum correction term of Eq.~\eqref{eq:donoghue_potential}, superposed linearly over the uniform mass distribution implicit in evaluating $\delta\Phi$ at the resonator center, is suppressed by a further, mass-independent factor of order $(\lP/\RH)^{2}\ln(\RH/\lP)\sim10^{-120}$. Both corrections are therefore utterly negligible over the entire range $\kIR\le\kappa\le\kP$ used in this work: the leading-order Newtonian potential used above is an excellent approximation throughout, and neither correction term of Eq.~\eqref{eq:donoghue_potential} needs to be included in Hypothesis~2.

\subsubsection{The self-consistent mode chain}

The resonator is deformed by every quantum added to it, so the allowed modes do not form the naive equally spaced sequence $\kappa_{n} = (n+1)\,\kIR$ set by the fundamental mode~\eqref{eq:kIR_def}. The physical content is a gravitational redshift: seen from outside, an excitation makes the resonator smaller and its modes oscillate more slowly, the fundamental mode included. Each mode is therefore displaced from its predecessor by its own self-gravitational shift,
\begin{equation}
  \kappa_{n+1} = \kappa_{n} + \delta \kappa(\kappa_{n}),
  \label{eq:recursion}
\end{equation}
with $\delta\kappa<0$, the redshift, and $\kappa_{0}=\kIR$ the lowest allowed mode, fixed by the causal boundary itself. The spacing along the chain is thus $|\delta\kappa| = \alpha\kappa^{2}$ and grows with $\kappa$, so that in the continuum limit, valid since $\alpha\kP\ll1$, $\mathrm{d}n/\mathrm{d}\kappa = 1/|\delta\kappa|$, which integrates with $\kappa(0)=\kIR$ to
\begin{equation}
  n(\kappa) = \frac{1}{\alpha}\left(\frac{1}{\kappa_{IR}} - \frac{1}{\kappa}\right).
  \label{eq:n_of_k}
\end{equation}
The mode density along the chain is therefore
\begin{equation}
\begin{aligned}
  \rho_{\text{chain}} (\kappa)=\frac{\mathrm{d}n}{\mathrm{d}\kappa} = & \frac{1}{\alpha \kappa^{2}}
  = \frac{\pi\,\kP^{2}}{3\,\kIR\,\kappa^{2}} = \frac{2c^{4}}{3GH_{0}\hbar\,\kappa^{2}}, \\&
  \qquad \kappa \geq \kIR.
\end{aligned}
  \label{eq:dn_dk}
\end{equation}
Two features of this result are essential. First, the $1/\kappa^{2}$ falloff is not postulated but follows directly from the self-gravitational deformation of the causal resonator. Second, there are no modes with $\kappa <  \kIR$: this is not a suppressed contribution but a genuine causal absence of degrees of freedom below the resonator's fundamental mode.
\subsection{Relation to the three-dimensional density of states}
\label{subsec:normalization}
For a field confined to a large three-dimensional volume $V$, the
standard momentum-space mode counting is
\begin{equation}
\mathrm{d}n_{\mathrm{3D}}
=
\frac{V}{(2\pi)^3}
\mathrm{d}^3\boldsymbol{\kappa}.
\label{eq:3d_mode_counting}
\end{equation}
Introducing spherical coordinates in momentum space gives
\begin{equation}
\mathrm{d}^3\boldsymbol{\kappa}
=
4\pi\kappa^2,\mathrm{d}\kappa.
\end{equation}
The standard radial density of states is therefore
\begin{equation}
\frac{\mathrm{d}n_{\mathrm{3D}}}{\mathrm{d}\kappa}
=
\frac{V}{2\pi^2}\kappa^2.
\label{n3D}
\end{equation}
The factor $\kappa^2$ originates purely from the angular degeneracy of
three-dimensional momentum space. It is a geometrical phase-space
factor and is independent of the dynamical mechanism determining
which of these states remain coherent in the causal resonator.
Accordingly, the
effective gravitational mode weighting is taken to have the form
\begin{equation}
W_H(\kappa)
=
\mathcal{N}_H
\frac{1}{\kappa^2},
\label{eq:WH_general}
\end{equation}
where $\mathcal{N}_H$ is a normalization constant with dimensions of
inverse length squared,
\begin{equation}
[\mathcal{N}_H]
=
\mathrm{m}^{-2}.
\end{equation}
Substitution into Eq.~\eqref{n3D}
gives an effective mode density $\rho_{\text{eff}} = \frac{dn_{\text{eff}}}{d \kappa}$
\begin{equation}
\rho_{\text{eff}}=\frac{\mathrm{d}n_{\text{eff}}}{\mathrm{d}\kappa}
=
\frac{V\mathcal{N}_H}{2\pi^2} = \frac{V\; W_H \kappa^2}{2\pi^2}.
\label{eq:rho_eff_3d}
\end{equation}
Thus, an inverse quadratic weighting of the three-dimensional
phase-space density compensates the ordinary $\kappa^2$ growth of the
radial phase-space measure. The resulting effective density per unit
wavenumber is independent of $\kappa$.
This observation should not be confused with the inverse quadratic
density of the self-gravitational chain itself. The two relations are
\begin{equation}
\rho_{\mathrm{chain}}(\kappa)
=
\frac{1}{\alpha\kappa^2},
\end{equation}
and
\begin{equation}
\frac{\mathrm{d}n_{\mathrm{3D}}}{\mathrm{d}\kappa}
=
\frac{V}{2\pi^2}\kappa^2.
\end{equation}
The first describes the spectral density generated by the
self-consistent recursion, whereas the second describes the geometric
degeneracy of three-dimensional momentum space. The causal-resonator
weighting connects these two descriptions only at the level of their
spectral dependence.
The normalization $\mathcal{N}_H$ cannot be obtained by directly
equating $\rho_{\mathrm{chain}}(\kappa)
=
\frac{\mathrm{d}n_{\mathrm{eff}}}{\mathrm{d}\kappa}$,
because the left-hand side counts the modes of a one-dimensional
self-consistent sequence, while the right-hand side includes the full
angular degeneracy of a three-dimensional quantum field. Such an
identification would implicitly assume a particular mapping between
the self-gravitating spectral chain and the angular structure of the
field modes, which has not been derived in the present framework.
The natural physical scale entering the normalization is the causal
infrared scale $\kappa_{\mathrm{IR}}$. It should be noted that the
dimension m$^{-2}$ alone does not select it: $\kappa_{\mathrm{IR}}\kP$
and $\kP^{2}$ are equally admissible on dimensional grounds and would
change the resulting vacuum-energy density by many orders of magnitude.
The choice is made on physical grounds, namely that the weighting
originates in the causal boundary, whose only scale is
$\kappa_{\mathrm{IR}}$, whereas $\kP$ enters the construction as a
limit of validity rather than as part of that boundary. The chain
density~\eqref{eq:dn_dk} does contain $\kP$ through $\alpha$, so this is
an argument about where the weighting originates and not a claim that no
other scale appears in the construction.
With this understanding, dimensional analysis
suggests
\begin{equation}
\mathcal{N}_H
=
K_{gr}
\kappa_{\mathrm{IR}}^2,
\label{eq:NH_Kgr}
\end{equation}
where $K_{gr}$ is a dimensionless coefficient. It collects two
contributions that the present work does not separate: the
normalization of the mapping between the self-gravitational mode chain
and the three-dimensional field spectrum, which is not derived and is
not bounded by unity, and the fraction of the gravitational vacuum
fluctuations that remains coherent against losses to other fields,
which is.
The effective causal-resonator weighting can consequently be written
as
\begin{equation}
W_H(\kappa)
=
K_{gr}
\frac{\kappa_{\mathrm{IR}}^2}{\kappa^2}.
\label{eq:wH_area}
\end{equation}
It can be concluded that
\begin{equation}
W_H(\kappa_{\mathrm{IR}})
=
K_{gr}  .
\end{equation}
The infrared scale enters because it is the only macroscopic
wavenumber characterizing the causal resonator. The spectrum begins at
\begin{equation}
\kappa
=
\kappa_{\mathrm{IR}},
\end{equation}
and no coherent causal-resonator modes exist below this value, $\kappa < \kappa_{\mathrm{IR}}$ because longer wavelengths would reach beyond the Hubble radius and would break causality.
Within the present approximation, the robust result of the
self-gravitational construction is the spectral dependence
\begin{equation}
W_H(\kappa)
\propto
\frac{\kappa_{\mathrm{IR}}^2}{\kappa^2},
\end{equation}
together with the causal lower bound $\kappa \geq\kappa_{\mathrm{IR}}$.
The dimensionless normalization coefficient $K_{gr}$ remains an
effective parameter of the mapping from the self-consistent
one-dimensional mode chain to the three-dimensional quantum field
spectrum.
The mode spectrum ends for different fields at different scales. There is no end for most quantum fields. Some fields reach symmetry breaking at a certain $\kappa$. For gravity the natural end is near the Planck scale.
\subsection{Lorentz Covariance of the Causal Resonator}
\label{subsec:lorentz_resonator}
\subsubsection{The causal resonator as an invariant structure}
\label{subsubsec:invariant_postulate}
The construction of Sec.~\ref{subsec:causal_resonator} rests on a
postulate that is analogous in status to the constancy of $c$
itself. In every inertial frame, only those modes are admitted that
respect causality within that frame's Hubble sphere: a mode whose
spatial extent would reach beyond the Hubble radius connects points
that are not causally related and is therefore excluded. Since the
phase velocity entering this construction is $c$ in every inertial
frame, for the gravitational as well as other fields,
the same causal boundary condition, and hence the same discrete mode
spectrum $\{\kappa_n\}$ generated by the self-gravitational
recursion, is obtained in every inertial frame. The causal resonator
is thus not a structure defined in one frame and subsequently
transformed into another; it is constructed independently and
identically in each frame from the causality requirement alone.
Spacetime itself, together with all fields propagating within it, is
embedded in this invariant causal resonator for each observer.
What is not invariant, and what must instead transform covariantly
under a boost, is the \emph{occupation} of this fixed mode structure
by a given physical excitation: the distribution of quanta among the
$\{\kappa_n\}$ that represents, for instance, a localized electron
or a light pulse. The physically meaningful question is therefore
not how the resonator basis transforms -- under the present
postulate it does not -- but how the occupation amplitudes of a wave
packet, expressed in the invariant basis, change between frames.
\subsubsection{Unitarity of the occupation transformation}
\label{subsubsec:unitarity}
Consider two inertial observers related by a Lorentz boost, each
with its own causal resonator carrying the same mode spectrum
$\{\kappa_n\}$. Let an arbitrary one-photon state be written as a
superposition of allowed resonator modes,
\begin{equation}
  |\Psi\rangle
  =
  \sum_n c_n\, |1,\kappa_n\rangle,
  \qquad
  \sum_n |c_n|^2 = 1.
\end{equation}
For a massless field,
\begin{equation}
  k_n^\mu k_{n\mu} = 0,
\end{equation}
and a Lorentz transformation maps each four-momentum according to
\begin{equation}
  k_n^\mu \longrightarrow k_n'^\mu = \Lambda^\mu{}_\nu k_n^\nu,
\end{equation}
so that the frequency associated with a given excitation transforms
according to the relativistic Doppler relation
\begin{equation}
  \omega_n' = \gamma\,\omega_n\left(1 - \beta\cos\theta_n\right).
\end{equation}
The frequency assigned to a particular excitation, and consequently
the energy assigned to the quantum by the observer, is therefore
frame-dependent, while the set of available modes is not: the
excitation is redistributed over the same invariant spectrum.
The state transformation itself is unitary,
\begin{equation}
  |\Psi'\rangle = U(\Lambda)\,|\Psi\rangle,
  \qquad
  U^\dagger(\Lambda)\,U(\Lambda) = \mathbb{I},
\end{equation}
so that
\begin{equation}
  \langle\Psi'|\Psi'\rangle
  = \langle\Psi|\,U^\dagger(\Lambda)U(\Lambda)\,|\Psi\rangle
  = \langle\Psi|\Psi\rangle
  = 1.
\end{equation}
A Lorentz boost therefore changes which modes of the invariant
spectrum are occupied, and with what amplitudes, but it does not
change the norm of the quantum state. The total probability and the
relative quantum weights are preserved, while the frequencies and
the total energy assigned to the state are Doppler shifted. For a
wave packet centered around a frequency $\nu_0$ one may write
schematically
\begin{equation}
  \langle E\rangle = h\nu_0,
  \qquad
  \langle E'\rangle = h\nu_0',
\end{equation}
with $\nu_0'$ related to $\nu_0$ by the appropriate Doppler factor.
The change in energy does not represent a loss of unitarity: energy
is the eigenvalue associated with the observer's time-translation
generator and is consequently frame-dependent, whereas the Hilbert
space norm is Lorentz invariant.
\subsubsection{The covariant spectral variable}
\label{subsubsec:covariant_variable}
The appropriate covariant variable for describing the occupation of
the invariant mode spectrum is the scalar product of the observer
four-velocity $u^\mu$ with the four-momentum $k^\mu$ of a massless
excitation. The frequency measured by an observer with four-velocity
$u^\mu$ is
\begin{equation}
  \omega_u(k) = \frac{|u_\mu k^\mu|}{\hbar},
\end{equation}
which in the observer's rest frame, $u^\mu = (c,0,0,0)$, reduces to
the usual photon frequency. Under a Lorentz transformation,
\begin{equation}
  u^\mu \to u'^\mu = \Lambda^\mu{}_\nu u^\nu,
  \qquad
  k^\mu \to k'^\mu = \Lambda^\mu{}_\nu k^\nu,
\end{equation}
the scalar product is invariant,
\begin{equation}
  u'_\mu k'^\mu = (\Lambda u)_\mu (\Lambda k)^\mu = u_\mu k^\mu,
\end{equation}
so that $\omega_{u'}(k') = \omega_u(k)$ when both the excitation and
the observer are transformed together. This should not be confused
with the Doppler transformation of the frequency measured by two
\emph{different} observers of the \emph{same} photon: if the same
photon is described with respect to a different observer, its
measured frequency changes according to the usual Doppler factor.
The invariant quantity is the observer--momentum scalar $u\cdot k$,
and this is the variable in terms of which the mode weighting
$W(k;u,\kc)$ of Sec.~\ref{subsec:planck} is formulated. The causal
resonator thus provides a physical origin for the object-relative
character of that weighting: the weighting is naturally expressed
relative to the observer whose causal resonator defines the mode
structure in which the interaction takes place.
\subsubsection{Displaced resonator centers}
\label{subsubsec:identical_basis}
For two comoving observers at different spatial locations, the
invariance stated above is complemented by an explicit
construction. Since the recursion coefficient
$\alpha = (3/\pi)\,\kIR/\kP^2$ depends only on $H_0$, which is by
the cosmological principle the same for every comoving observer
regardless of location, the sequence $\{\kappa_n\}$ is identical for
any two such observers, independent of the spatial separation
between their resonator centers. The deformation
$\delta\kappa = -\alpha\kappa^2$ determines \emph{which} wavenumbers
are admitted into the spectrum; it does not depend on the location
of the resonator origin.
Crucially, the causal boundary constrains only the admissible
wavenumbers, not the phases of the corresponding modes. As discussed
in Sec.~\ref{subsec:causal_resonator}, the restriction of the
spectrum arises from the requirement that the basis permit a unique
assignment of occupation numbers on the closed causal path, rather
than from a reflecting boundary imposing
nodes on the field amplitude. No physical mechanism enforces
vanishing amplitude at the Hubble radius; the horizon is a causal
limit, not a reflecting wall. The mode phases are therefore not
fixed by the construction and remain free parameters of the
excitation.
This has a direct consequence for observers with displaced
resonator centers. A rigid spatial displacement by $\bm{d}$ acts on
each mode of the shared spectrum purely as a phase shift,
\begin{equation}
  \varphi_n \;\longrightarrow\; \varphi_n + \bm{\kappa}_n\cdot\bm{d},
  \label{eq:phase_shift_translation}
\end{equation}
leaving the set $\{\kappa_n\}$ and the magnitudes $|c_n|$ of the
occupation amplitudes unchanged. A wave packet
\begin{equation}
  |\Psi\rangle = \sum_n c_n\,|1,\kappa_n\rangle,
  \qquad
  c_n = |c_n|\,e^{i\varphi_n},
\end{equation}
described by one comoving observer is therefore described by any
other comoving observer as the same superposition of the same modes,
with phases shifted according
to~\eqref{eq:phase_shift_translation}. The relative phases encode
the spatial position of the packet, so that the packet is correctly
localized at the displaced position in the second description
without requiring any change of basis. Translation between
comoving causal resonators is thus realized as a pure phase
transformation, which manifestly preserves both the mode spectrum
and the norm $\sum_n|c_n|^2$.
\subsubsection{Composition of translations and boosts}
\label{subsubsec:composition}
A general transformation between an observer at rest at one location
and a boosted observer at a different location decomposes into a
translation to a comoving observer at the second location, treated
in Sec.~\ref{subsubsec:identical_basis}, followed by a local boost
at that location, treated in Sec.~\ref{subsubsec:unitarity}. Both
steps act on the occupation of the same invariant mode spectrum: the
translation as a pure phase transformation, the boost as a unitary
redistribution of the occupation amplitudes accompanied by the usual
Doppler shift of the frequencies assigned to the excitation. Since
Lorentz transformations compose, and since neither step alters the
admissible spectrum $\{\kappa_n\}$ or the norm of the state, the
composite transformation preserves both properties as well. A photon
or electron wave packet can therefore be consistently followed
through an arbitrary sequence of spatial translations and local
boosts between causal resonators, with its position encoded in the
mode phases and its energy content in the occupation amplitudes.
\subsubsection{The continuum limit and compatibility with
               established physics}
\label{subsubsec:compatibility_current}
The ordinary, continuous Lorentz transformation of occupation
amplitudes, familiar from standard quantum field theory, is
recovered here as an approximation. Because the causal resonator
carries an extraordinarily large number of modes, the discreteness
of the underlying spectrum is not resolvable in practice, and the
usual continuous Doppler redistribution of a wave packet's
occupation is obtained to excellent accuracy. An electron at rest in
one frame acquires higher-frequency excitations in its boosted
description, and a boosted light pulse is blueshifted, exactly as
expected from the ordinary Doppler relation applied to the
occupation of the invariant resonator modes.
For every regime accessible to measurement, this description
reproduces established physics. As shown in
Sec.~\ref{subsec:loss_origin}, the causal infrared scale $\kIR$
is smaller than the local decoherence scale $\kc$ governing
matter-field physics by many orders of magnitude, for any
experimentally realized boost, including those achieved in particle
accelerators and inferred from ultra-high-energy cosmic rays. Local,
decoherence-dominated processes -- the anomalous magnetic moment,
the Lamb shift, and the quantum-electrodynamic observables of
Ref.~\cite{Rembe2026} -- are governed by $\kc$ alone, and the causal
resonator introduces no measurable modification to their known
Lorentz covariance. Whether the invariance postulated here could in
principle be distinguished observationally from alternative
assumptions about the transformation behavior of $\kIR$ is a
question of new physics beyond currently available data, and is
addressed as an open direction in Sec.~\ref{sec:discussion}.
\section{Ultraviolet power counting at one loop}
\label{sec:powercounting}
\subsection{Euclidean loop integral with mode weighting}
\label{subsec:euclidean}
The effective mode structure is a property of the graviton mode, not of the loop integration, and therefore enters once for each internal graviton line: for internal momenta $k_{1},\dots,k_{I}$ the weighting appears as the product $\prod_{j}W(k_{j})$. This is the implementation used in the consistency argument of Sec.~\ref{sec:bianchi}, where the weighting multiplies each internal propagator, $D_{\mu\nu\rho\sigma}(k)\to W(k)D_{\mu\nu\rho\sigma}(k)$, and it is the only one that is well defined. A loop momentum is a routing variable: attaching the weighting to it would make the result depend on which internal line the integration variable is assigned to, whereas attaching it to each line is invariant under the shifts that relate equivalent routings. Since the weighted integrals converge, such shifts are legitimate, which is what the telescoping argument of Sec.~\ref{sec:bianchi} requires.
After Wick rotation to Euclidean momentum space a generic one-loop contribution takes the form
\begin{equation}
  I_{1\text{-loop}}
  \sim
  \int \frac{\mathrm{d}^{4} \KE}{(2\pi)^{4}}\;
  \prod_{j=1}^{I} W(k_{j};\, \kc)\;
  F(\KE,\, \text{external}),
  \label{eq:loop_integral}
\end{equation}
where $\KE$ denotes the magnitude of the Euclidean loop momentum, $I$ the number of internal graviton lines, $F$ the standard graviton propagator and vertex structure of the diagram, and $W$ the effective object-relative weighting of Eq.~\eqref{eq:weighting_restframe} below. In the ultraviolet all internal momenta of a one-loop graph grow together with $\KE$, so the product reduces to $W(\KE;\kc)^{I}$, and with $\mathrm{d}^{4}\KE = 2\pi^{2}\KE^{3}\,\mathrm{d}\KE$ the radial form is
\begin{equation}
  I_{1\text{-loop}}
  \sim
  \frac{1}{8\pi^{2}}
  \int \mathrm{d}\KE\;
  \KE^{3}\, W(\KE;\, \kc)^{I}\; F .
  \label{eq:loop_radial}
\end{equation}
\subsection{Weighting to internal graviton lines}
\label{subsec:transfer}
In Ref.~\cite{Rembe2026} an object-relative ultraviolet weighting was introduced for internal electromagnetic modes. Its motivation is that a sufficiently localized field mode carries energy $E=\hbar\omega$ and that this energy gravitates, so that with $\kappa=|\bm{k}|=\omega/c$ the weak-field self-backreaction shifts the wavenumber by
\begin{equation}
  \delta \kappa \sim -\,\frac{\kappa^{3}}{\kP^{2}},
  \qquad
  \kP = \sqrt{\frac{c^{3}}{\hbar G}} .
  \label{eq:freq_shift}
\end{equation}
The argument depends only on the energy carried by the mode and on its localization scale, not on the spin structure of the field, and therefore applies to graviton modes without modification.
Because the mode structure is object-relative, the weighting appropriate to a calculation is fixed by the process, and two cases must be kept apart. Loop corrections describe \emph{local} interactions: the participating fields are localized by the interaction itself, coherence does not extend beyond that localization, and $\kc$ is correspondingly a local scale. That is the case treated here. The globally coherent vacuum of the causal resonator is governed by $\kIR$ instead and is treated in Sec.~\ref{sec:darkenergy}; the relation between the two regimes is discussed in Sec.~\ref{subsec:loss_origin}. The weighting $W_{\mathrm{H}}$ of Eq.~\eqref{eq:WH_final} is therefore \emph{not} the weighting entering the loop integrals of this section, and conversely. That the same graviton field is counted differently in the two cases is a consequence of the object-relative postulate rather than an inconsistency to be removed: the effective mode content is defined relative to the localization that defines the process, and the two processes are localized differently. It remains an assumption, and a load-bearing one, since the all-orders criterion of Sec.~\ref{subsec:goroff_sagnotti} uses the local exponent and not the one obtained for the fixed resonator in Sec.~\ref{subsec:causal_resonator}.
\subsection{Lorentz consistency}
\label{subsec:lorentz}
The weighting, and with it the scale at which the mode content is suppressed, is formulated as a function of the Lorentz scalar $|u \cdot k|$ rather than of the laboratory three-momentum, so that it is preserved under $u^{\mu}\to\Lambda^{\mu}{}_{\nu}u^{\nu}$, $k^{\mu}\to\Lambda^{\mu}{}_{\nu}k^{\nu}$. The same effective mode structure is therefore realized in every inertial frame once the graviton mode is described relative to the localized interaction object and its invariant scale $\kc$. For real gravitons $k^{\mu}k_{\mu}=0$, so a condition formulated in terms of $k^{\mu}k_{\mu}$ alone could not distinguish low from high object-relative frequency; this is the same reason as in the electromagnetic case for using $u\cdot k$.
\subsection{The characteristic scale and the suppression exponent}
\label{subsec:planck}
In the QED application of Ref.~\cite{Rembe2026} the scale $\kc$ was fixed by matching the weighted one-loop anomalous magnetic moment to the Schwinger value, which yielded a scale near the electron Compton wavelength; there the localization is supplied by the interacting matter field. For a purely gravitational loop no such external localization scale exists. The only scale available is the one gravity itself provides, and the unique combination of $G$, $\hbar$ and $c$ with the dimension of a wavenumber is $\kP=1/\lP=\sqrt{c^{3}/\hbar G}$, so that
\begin{equation}
  \kc = \kP
  \label{eq:kc_planck}
\end{equation}
without free parameters and without a matching condition. This is an argument from the absence of an external localization scale, not from the universality of the coupling: a universal coupling constant does not by itself make the localization universal, and a graviton exchanged between two localized objects is localized on their scale rather than on $\lP$.
At $\kappa\sim\kP$ the dimensionless graviton self-coupling of Eq.~\eqref{eq:alpha_g_def} becomes of order unity and two-loop graviton--graviton contributions become comparable to the one-loop terms retained here. These are interactions of a mode with further graviton degrees of freedom, so coherence begins to be destroyed at this scale (Sec.~\ref{subsec:loss_origin}). A decohered mode is localized on its own scale, and for such a mode Eq.~\eqref{eq:freq_shift} applies with $R\sim1/\kappa$; carrying the recursion of Sec.~\ref{subsec:causal_resonator} through with $\mathrm{d}n/\mathrm{d}\kappa=1/|\delta\kappa|$ gives the falloff $\kP^{3}/\kappa^{3}$ obtained in Ref.~\cite{Rembe2026}. Writing the exponent as $s$, the effective mode structure for internal graviton lines is
\begin{equation}
  W(k;\, u,\, \kP)
  =
  \begin{cases}
    1, & |u \cdot k| \leq c\, \kP, \\[4pt]
    \left(\dfrac{c\,\kP}{|u \cdot k|}\right)^{\!s}, & |u \cdot k| > c\, \kP,
  \end{cases}
  \label{eq:weighting_planck}
\end{equation}
and in the rest frame of the interaction object
\begin{equation}
  W(\KE;\, \kP)
  =
  \begin{cases}
    1, & \KE \leq \kP, \\[4pt]
    \left(\dfrac{\kP}{\KE}\right)^{\!s}, & \KE > \kP .
  \end{cases}
  \label{eq:weighting_restframe}
\end{equation}
The leading value is $s=3$. Reference~\cite{Rembe2026} motivates it in two independent ways that agree: the self-backreaction estimate, in which $\delta\omega/\omega\sim\Phi/c^{2}$ together with $E=\hbar\omega$ and $R\sim1/\kappa$ produces a spacing $\propto\kappa^{3}$; and a three-dimensional overlap argument, in which the number of subvolumes a mode resolves in a localized interaction region grows as $\kappa^{3}$ up to $\kc$. That reference explicitly does not claim to derive the weighting from the self-backreaction estimate alone. The resonator case of Sec.~\ref{subsec:causal_resonator} is more tightly constrained and yields a different exponent: there the localization scale is the fixed $\RH$ rather than the mode's own wavelength, and $\Phi$ is evaluated explicitly at the centre of a uniform sphere, which is why it obtains $2$ rather than $3$. In either case $s$ is a property of the mode and not of the loop measure, so it is the same at every loop order and for every diagram, as the per-line implementation requires.
Below $\kP$ the construction operates inside its domain of validity. Above it, Eq.~\eqref{eq:freq_shift} is a weak-field estimate whereas $\alpha_g\gtrsim1$ holds there by construction, so $s=3$ indicates the leading behaviour rather than establishing an exact value. The first correction can be evaluated explicitly and is informative. In the configuration relevant here, $M=\hbar\kappa/c$ and $r\sim1/\kappa$, both correction terms of Eq.~\eqref{eq:donoghue_potential} are of relative order $\alpha_g$ itself, $3GM/rc^{2}=3\alpha_g$ and $(41/10\pi)G\hbar/r^{2}c^{3}=(41/10\pi)\alpha_g$, so that they are not an independent refinement but the expansion in $\alpha_g$,
\begin{equation}
  \frac{\Phi}{c^{2}} = -\,\alpha_g\left(1+c_{1}\,\alpha_g+\dots\right),
  \; \;
  c_{1}=3+\frac{41}{10\pi}\simeq 4.31 .
  \label{eq:phi_corrected}
\end{equation}
This differs from the situation in Hypothesis~2, where the same two corrections were shown in Sec.~\ref{subsec:causal_resonator} to be negligible because there $r=\RH$ and the two differ by $61$ orders of magnitude. Repeating the recursion with the corrected shift gives
\begin{equation}
  W(\kappa) \;\propto\; \frac{1}{x^{3}\left(1+c_{1}x^{2}\right)},
  \qquad x=\frac{\kappa}{\kP},
  \label{eq:W_corrected}
\end{equation}
whose asymptotic falloff is $s_{\mathrm{eff}}=5$; each further order of the expansion steepens it by two further powers, so long as the coefficients remain positive, as the two known ones are.
Two reservations attach to this. The truncation is not controlled: for $x>1$ the retained correction dominates, so $s_{\mathrm{eff}}=5$ indicates a direction rather than a value. And Eq.~\eqref{eq:W_corrected} does not match continuously onto the lower branch unless renormalized, since $W(1)\propto1/(1+c_{1})\simeq0.19$; that factor of $5.3$ is itself the statement that the expansion is not small at $\kP$. What is robust is the sign: both known coefficients are positive, so the corrections strengthen the suppression, and $s\ge3$ follows independently of their values. Only this one-sided bound is used below. Independently of the expansion, a mode of wavenumber $\kappa$ lies within its own Schwarzschild radius once $\kappa>\kP/\sqrt{2}$, which suggests that a sharp termination may be closer to the physical situation than any power law; the results of Sec.~\ref{subsec:selfenergy} are bracketed by these possibilities.
For interactions on cosmological scales the same object-relative reasoning identifies the second natural scale $\kIR$, set by $\lambda_0=4\RH$ (Eq.~\eqref{eq:kIR_def}). Unlike $\kP$ it is not determined by $G$, $\hbar$ and $c$ alone but requires the observed $H_{0}$; it is nevertheless free of adjustable parameters.
\subsection{One-loop finiteness of the graviton self-energy}
\label{subsec:selfenergy}
The one-loop graviton self-energy is the most direct test case. Its unweighted integrand behaves as $F_{\Sigma}(\KE)\sim\KE^{-2}$, the numerator $\KE^{2}$ arising from the gravitational vertex and the denominator $\KE^{4}$ from the two graviton propagators, so that the unweighted radial integral $\int^{\infty}\mathrm{d}\KE\,\KE$ diverges linearly, reproducing the divergence identified by 't\,Hooft and Veltman~\cite{tHooft1974}. That divergence arises entirely from mode content above $\kP$. The diagram carries $I=2$ internal graviton lines, so the weighting enters squared. The region below $\kP$ contributes $\int_{0}^{\kP}\mathrm{d}\KE\,\KE=\tfrac{1}{2}\kP^{2}$, finite because the region is bounded, and the region above contributes
\begin{equation}
  I_{\Sigma}^{\,\KE>\kP}
  \sim \kP^{2s}\int_{\kP}^{\infty} \mathrm{d}\KE\;\KE^{1-2s}
  = \frac{\kP^{2}}{2s-2},
  \label{eq:selfenergy_uv_tail}
\end{equation}
which converges for every $s>1$. The one-loop graviton self-energy is therefore ultraviolet finite,
\begin{equation}
  I_{\Sigma} \sim \left(\frac{1}{2}+\frac{1}{2s-2}\right)\kP^{2},
  \label{eq:selfenergy_total}
\end{equation}
of order $\kP^{2}$ and set entirely by the Planck scale. The leading value $s=3$ gives $\tfrac{3}{4}\kP^{2}$; the corrected weighting~\eqref{eq:W_corrected}, matched to $W(\kP)=1$, gives $0.64\,\kP^{2}$; and $s\to\infty$, a sharp termination at $\kP$, gives $\tfrac{1}{2}\kP^{2}$. The three readings bracket the result between $\tfrac{1}{2}\kP^{2}$ and $\tfrac{3}{4}\kP^{2}$. With the corrected weighting, $79\%$ of Eq.~\eqref{eq:selfenergy_total} originates below $\kP$ and $99\%$ below $1.5\,\kP$, so the region in which $\alpha_g>1$ contributes a minor part, although the integrand still peaks near $\kappa\simeq\kP$.
\subsection{General one-loop power counting}
\label{subsec:general}
The argument extends to the general class of one-loop graviton diagrams. Let $\mathcal{D}$ denote the superficial degree of divergence of the unweighted diagram, so that its radial integrand behaves as $\KE^{\mathcal{D}-1}$; for the self-energy, $\KE^{3}\cdot\KE^{-2}$ gives $\mathcal{D}=2$. The bounded region then contributes $\kP^{\mathcal{D}}/\mathcal{D}$ for $\mathcal{D}>0$, and above $\kP$ the integrand is $\kP^{sI}\KE^{\mathcal{D}-1-sI}$, so that
\begin{equation}
  I^{\,\KE>\kP}
  \sim \kP^{sI}\int_{\kP}^{\infty} \mathrm{d}\KE\; \KE^{\mathcal{D}-1-sI}
  = \frac{\kP^{\mathcal{D}}}{\,sI-\mathcal{D}\,},
  \label{eq:general_uv_tail}
\end{equation}
which converges precisely under
\begin{equation}
  s\,I > \mathcal{D} .
  \label{eq:p_condition}
\end{equation}
Here $I$ is again the number of internal graviton lines. For the self-energy, $\mathcal{D}=2$ and $I=2$, this is $s>1$, comfortably met by $s=3$; diagrams with more internal lines or a more strongly falling kernel relax it further, and for $\mathcal{D}\le0$ the ultraviolet region converges for any positive $s$. Only internal graviton lines are weighted here, as only internal photon lines are in Ref.~\cite{Rembe2026}; if the mode structure is a property of the causal resonator rather than of a particular field, the internal lines of every field carry it, each with its own characteristic scale, and the conditions obtained here are then lower bounds, since further factors raise $sI$ and can only improve convergence. Under Eq.~\eqref{eq:p_condition} no contribution of the class considered here diverges in the ultraviolet, and no divergent counterterms are required at one loop. Writing the condition in terms of $\mathcal{D}$ rather than of the kernel exponent makes it identical in form to the all-orders condition of Sec.~\ref{subsec:goroff_sagnotti}, where $\mathcal{D}$ is replaced by the $L$-loop degree $\mathcal{D}_{L}$.

\subsection{Scope of the finiteness result}
\label{subsec:finiteness_scope}
What carries this result is the threshold~\eqref{eq:p_condition}, not a particular value of $s$ and not an externally imposed cutoff. The construction supplies the scale $\kP$ at which the graviton self-coupling reaches order unity and a dissipative channel opens, and any suppression exceeding $\mathcal{D}/I$ suffices; at one loop the threshold lies far below the derived value $s=3$, so the conclusion is insensitive to the exponent, and it is met by a wide class of behaviours including a sharp termination of the spectrum. In this sense the result differs from a regularization: the scale is not chosen to make the integrals converge, and the outcome does not depend on the detailed form of the suppression.
This bears directly on the counterterms. In the effective field theory treatment of gravity~\cite{Donoghue1994, Wilson1974, Weinberg1979} the cutoff is an external parameter, the higher-curvature operators generated at successive loop orders carry divergent coefficients, and each must be absorbed into a separately measured coupling; that is where predictivity is lost. Here the cutoff follows from the construction and the integrals converge, and the criterion of Sec.~\ref{subsec:goroff_sagnotti} covers the two-loop divergence of pure gravity found by Goroff and Sagnotti~\cite{Goroff1986}: the same bound applies to every internal graviton line and hence to both loop momenta, and every two-loop topology with more than one vertex satisfies $L+3V>5$ with a wide margin. This holds under the premise that the weighting attaches to the lines entering the nonlinear vertices, and it leaves out the single-vertex class identified there. The curvature-cubed operator is still generated, but with a finite coefficient that the mode structure computes rather than a divergent one that must be absorbed. Counterterms are in this sense not required, and the loss of predictivity that follows from needing them does not arise in the same way.

Two qualifications attach to that statement, and they should not be conflated. The computed coefficients depend on the mode content above $\kP$, which is not derived here; the dependence is weak, since the weighted integrand falls steeply beyond $\kP$, but it is not zero. And the construction does not determine which operators the \emph{underlying} action contains, because the Einstein--Hilbert form is assumed rather than derived — a limitation it shares with every effective treatment, and one that concerns the starting point rather than the loop expansion. The infrared predictions of the effective theory are untouched (Sec.~\ref{subsec:infrared}).
A limitation of the framework as a whole should be recorded here. The characteristic scale $\kc$ is object-relative and is assigned in each application on physical grounds: the Compton scale for the anomalous magnetic moment and the atomic scale for the Lamb-shift estimate in Ref.~\cite{Rembe2026}, $\kP$ for the graviton loops of this section, and $\kIR$ for the coherent cosmological vacuum of Sec.~\ref{sec:darkenergy}. Each assignment is physically motivated and each yields a scale of the expected magnitude, but none is computed from a dynamical criterion. A derivation of the decoherence lengths that determine $\kc$ is the principal missing element of the framework, in the loop sector as much as in the cosmological one.

\subsection{Infrared sector unaffected}
\label{subsec:infrared}
Since $W(\KE;\,\kP)=1$ for $\KE\le\kP$, the weighting leaves the infrared part of every loop integral unchanged. The quantum corrections to the Newtonian potential~\eqref{eq:newtonian_corrections}, whose coefficients are determined by the infrared structure of the one-loop integrals at $\KE\ll\kP$, are therefore unaffected, the long-range predictions of the effective field theory of gravity~\cite{Donoghue1994} are fully preserved, and the Newtonian limit is recovered without modification.
\section{Causal-resonator vacuum energy and emergence of the
         dark-energy scale}
\label{sec:darkenergy}
\subsection{Finite vacuum energy from the modified mode density}
\label{subsec:vacuumenergy}
In the standard treatment of quantum field theory on flat spacetime,
the vacuum energy density diverges as
\begin{equation}
  \rho_{\mathrm{vac}}^{\mathrm{std}}
  \sim \int_{0}^{\infty} \mathrm{d}^{3}\bm{\kappa}\;
       \hbar c\,\kappa
  \;\longrightarrow\; \infty.
  \label{eq:standard_vacuum}
\end{equation}
This divergence is the origin of the cosmological constant problem:
if interpreted as a gravitational source on the right-hand side of
the Einstein equation, the resulting energy density exceeds the
observed dark energy density by approximately 120 orders of
magnitude, even if a cutoff is imposed at the Planck scale instead
of infinity.
With the causal-resonator mode structure of
Sec.~\ref{sec:weighting}, the effective vacuum energy density
becomes finite. Two features of that construction are essential
here. First, no modes exist below $\kIR$: this is not a suppressed
contribution but a genuine causal absence of degrees of freedom.
Second, the effective mode weighting falls as $\kappa^{-2}$ above
$\kIR$, so that the integrand grows only linearly in $\kappa$ rather
than cubically.

\subsection{Geometric interpretation and equation of state}
\label{subsec:geometric}
The gravitational self-backreaction of the gravitational vacuum
modes does not contribute as a matter source $T_{\mu\nu}$ on the
right-hand side of the linearized quantized Einstein
equation~\eqref{eq:quantized_field}. Rather, the modification of the
effective mode structure alters the geometry directly, so that the
resulting finite vacuum energy appears as a geometric term on the
left-hand side, without requiring a separate cosmological constant in
addition to matter. Which side of the equation the term is written on
fixes only a relative sign, however, and does not by itself decide
whether the contribution accelerates or decelerates the expansion.
That question is settled by the tensor structure of the contribution,
which the construction determines as follows.

The density obtained below, Eq.~\eqref{eq:rho_closed}, is assembled
from $\kIR$, $\kP$ and the coefficient $K_{gr}$. Both wave numbers are
Lorentz scalars: $\kIR=\pi H/2c$ is fixed by the expansion rate, and
$\kP$ by $\hbar$, $c$ and $G$. By the postulate of
Sec.~\ref{subsubsec:invariant_postulate} the resonator is not
transported from one frame into another but constructed identically in
each from the causality requirement alone, so that the same spectrum
and the same weighting are obtained in every inertial frame. Every
inertial observer therefore assigns the unexcited resonator the same
energy density $\rho_{\Lambda}$. This is not an additional assumption:
it is required for the consistency of the construction, since
otherwise the expansion rate inferred from the vacuum contribution
would depend on the observer.

Let $u^{\mu}$, normalized to $u^{\mu}u_{\mu}=1$, be the four-velocity
of an arbitrary inertial observer; the energy density that observer
measures is $T_{\mu\nu}u^{\mu}u^{\nu}$. If this quantity equals
$\rho_{\Lambda}c^{2}$ for every timelike $u^{\mu}$, then the difference
$T_{\mu\nu}-\rho_{\Lambda}c^{2}g_{\mu\nu}$ has vanishing quadratic form
in every timelike direction, and a symmetric tensor whose quadratic form
vanishes on an open set of directions vanishes identically. Hence
\begin{equation}
  T_{\mu\nu} = \rho_{\Lambda}\,c^{2}\,g_{\mu\nu},
  \qquad
  \mathcal{P} = -\rho_{\Lambda}c^{2},
  \label{eq:eos_from_invariance}
\end{equation}
so that the equation of state is not assigned but follows from the
frame-independence of the resonator. With $\rho_{\Lambda}>0$ the
acceleration equation~\cite{Misner1973} gives
$\ddot{a}/a=-(4\pi G/3)(\rho_{\Lambda}+3\mathcal{P}/c^{2})
=+(8\pi G/3)\rho_{\Lambda}>0$: the contribution is repulsive and drives
de Sitter expansion. The pressure term does not merely cancel the
attraction associated with $\rho_{\Lambda}$ -- that case would
correspond to $\mathcal{P}=-\rho_{\Lambda}c^{2}/3$ -- but reverses it,
with twice the magnitude.

It is necessary to state why the radiation-like value does not apply
here instead. Written as a sum over on-shell modes, a vacuum stress
tensor would read
\begin{equation}
  T_{\mu\nu} \;=\; \sum_{n} W_{n}\,
  \frac{\hbar\,c^{2}}{2}\,
  \frac{k_{\mu}^{(n)}k_{\nu}^{(n)}}{\omega_{n}},
  \label{eq:onshell_sum}
\end{equation}
whose trace vanishes identically, since $k^{\mu}k_{\mu}=0$ for every
massless mode and a positive weighting $W_{n}$ cannot alter this, so
that $\rho=3\mathcal{P}$ follows irrespective of the weighting profile.
This form presupposes that the modes are ordinary momentum eigenstates
transforming as null four-vectors. No such set can be boost invariant:
a boost blueshifts forward-propagating and redshifts
backward-propagating modes and thus maps the set onto a different,
anisotropic one, and the energy density it yields is correspondingly
frame-dependent. This is precisely why a momentum cutoff violates
Lorentz invariance in the standard treatment, and it is why
Eq.~\eqref{eq:onshell_sum} cannot describe the unexcited resonator,
whose density is the same for every observer.

The construction proposed here does not transform the resonator in
this way. The causal resonator is re-defined independently in every
inertial frame from the causality requirement, as stated in
Sec.~\ref{subsubsec:invariant_postulate}. Four-momentum is carried by
the excitations of the resonator, which transform in the ordinary way
and are redistributed over the spectrum according to the Doppler
relation (Sec.~\ref{subsubsec:unitarity}), and not by the basis
itself; physical content is accordingly always accessed through the
excitations. Lorentz invariance is required of this excitation sector,
where it is preserved by construction, and
Eq.~\eqref{eq:onshell_sum} is the stress tensor of a population of
on-shell quanta and therefore describes excitations. The unexcited
resonator carries no such population, and its contribution is fixed
instead by Eq.~\eqref{eq:eos_from_invariance}.

The three-dimensional counting measure $(V/2\pi^{2})\kappa^{2}$ used in
Sec.~\ref{subsec:normalization} is not transported between frames
either. It is applied in each frame to the same scalar interval
$[\kIR,\kP]$ by the same prescription, so that it counts the same
spectrum for every observer, and the resulting
density~\eqref{eq:rho_closed} accordingly contains Lorentz scalars
alone. What the measure does not settle is its own normalization
relative to the one-dimensional mode chain; that is the open question
$\mathcal{N}_{H}$ of Sec.~\ref{subsec:normalization}, and it concerns
the coefficient rather than the frame-independence relied on here.

\subsection{Order-of-magnitude estimate}
\label{subsec:estimate}
The effective gravitational vacuum energy density is obtained using
the standard three-dimensional mode-counting density of states,
$\mathrm{d}n = V\,\mathrm{d}^{3}\bm{\kappa}/(2\pi)^{3}$, consistent
with the normalization used for the Euclidean loop integrals of
Sec.~\ref{sec:powercounting}, together with the causal-resonator
weighting
\begin{equation}
  W_{\mathrm{H}}(\kappa)
  = K_{gr}\,\frac{\kIR^{2}}{\kappa^{2}},
  \qquad \kIR \leq \kappa \leq \kP,
  \label{eq:WH_final}
\end{equation}
introduced in Sec.~\ref{sec:weighting}. Summing over the two
physical graviton polarizations, each of which carries its own
independent mode spectrum, and assigning each mode the zero-point
energy $\hbar\omega/2$, the vacuum energy density is
\begin{equation}
  \rho_{\Lambda}
  = 2\int_{\kIR}^{\kP}
    \frac{\kappa^{2}\,\mathrm{d}\kappa}{2\pi^{2}}\;
    \frac{\hbar\kappa}{2c}\;
    W_{\mathrm{H}}(\kappa)
  = K_{gr}\,\frac{\hbar\,\kIR^{2}\,\kP^{2}}{4\pi^{2}c},
  \label{eq:rho_integral}
\end{equation}
where terms of order $(\kIR/\kP)^{2}\sim10^{-122}$ have been
dropped. Inserting $\kIR = \pi H_{0}/(2c)$ and
$\kP^{2} = c^{3}/(\hbar G)$ gives the closed form
\begin{equation}
  \rho_{\Lambda} = K_{gr}\,\frac{H_{0}^{2}}{16G}.
  \label{eq:rho_closed}
\end{equation}
The upper limit of the integral is set by the Planck wave number. The
reason is not a breakdown of the self-energy estimate of Hypothesis~2:
within that fixed-resonator construction, both the classical and the
quantum correction to the potential remain entirely negligible over the
whole range $\kIR\le\kappa\le\kP$
(Sec.~\ref{subsec:causal_resonator}). The relevant limitation is more
basic and independent of any resonator geometry. Each graviton vertex
obtained from the nonlinear expansion of the Einstein--Hilbert action
carries the coupling $\kappa_{\mathrm g}=\sqrt{16\pi G/c^{4}}$, so that
the dimensionless expansion parameter controlling graviton
self-interaction at wave number $\kappa$ is
\begin{equation}
  \alpha_g(\kappa) = \frac{G\hbar\kappa^{2}}{c^{3}}
  = \left(\frac{\kappa}{\kP}\right)^{2}.
  \label{eq:alpha_g_def}
\end{equation}
This is the same combination that fixes the characteristic scale
$\kc\approx\kP$ in Sec.~\ref{subsec:planck} and the same parameter that
suppresses higher-loop graviton self-interactions relative to the
one-loop terms retained here (Sec.~\ref{subsec:goroff_sagnotti}). It
reaches order unity at $\kappa\sim\kP$ irrespective of the
configuration in which a mode is considered. There, two-loop
graviton--graviton contributions of the type identified by Goroff and
Sagnotti~\cite{Goroff1986} become comparable to the one-loop
contributions on which the present mode counting rests, and the
linearized description requires an ultraviolet completion. The Planck
wave number is therefore an effective limit of validity rather than a
sharply characterized crossover, and it is not a derived cutoff.
Because the integrand in~\eqref{eq:rho_integral} grows only linearly in
$\kappa$, the result is dominated by modes near the upper limit and
scales as the square of the wave number at which the coherent spectrum
ends. Equation~\eqref{eq:rho_closed} therefore uses $\kP$ as a
reference value; the dependence on the actual endpoint is carried
explicitly in Sec.~\ref{subsec:desitter_fixedpoint}, where it absorbs
the residual coefficient. What the result does not depend on is the
mode structure beyond that endpoint, since decohered modes do not enter
the coherent sum at all.
With $K_{gr} = 1$, Eq.~\eqref{eq:rho_closed} yields
\begin{equation}
  \rho_{\Lambda}(K_{gr}=1)
  = \frac{H_{0}^{2}}{16G}
  \approx 4.5\times10^{-27}\,\mathrm{kg\,m^{-3}},
  \label{eq:rho_Kgr1}
\end{equation}
corresponding to
$\Omega_{\Lambda} = \rho_{\Lambda}/\rho_{\mathrm{crit}} \approx 0.52$
with $\rho_{\mathrm{crit}} = 3H_{0}^{2}/(8\pi G)$. The observed dark
energy density inferred from the Planck 2018 cosmological
parameters~\cite{Planck2020} is
\begin{equation}
  \rho_{\Lambda,\mathrm{obs}}
  \approx 5.84\times10^{-27}\,\mathrm{kg\,m^{-3}},
  \qquad
  \Omega_{\Lambda,\mathrm{obs}} \approx 0.68.
  \label{eq:rho_observed}
\end{equation}
The estimate therefore reproduces the correct sign and the correct
order of magnitude of the observed dark energy density from the two
parameter-free scales $\kIR$ and $\kP$ together with the number of
physical graviton polarizations alone, falling short of the observed
value by a factor of approximately $1.3$. Requiring exact agreement fixes
the dimensionless normalization coefficient introduced in
Sec.~\ref{sec:weighting} to
\begin{equation}
  K_{gr}
  = \frac{16G\,\rho_{\Lambda,\mathrm{obs}}}{H_{0}^{2}}
  \approx 1.31 .
  \label{eq:Kgr_value}
\end{equation}
That the required value is of order unity rather than differing from
unity by many orders of magnitude is the nontrivial property of the
construction, and it is what the following two subsections address:
where the coefficient comes from, and why it is not to be read off at
the present epoch.

\subsection{Normalization at the de Sitter fixed point}
\label{subsec:desitter_fixedpoint}
Equation~\eqref{eq:Kgr_value} evaluates $\kIR$ with the present Hubble
rate. This is the natural choice only if the causal resonator follows
the instantaneous Hubble radius, and that behaviour is excluded on
cosmological grounds (Sec.~\ref{subsec:scope2}): the mode basis must be
frozen, and one must then state at which causal scale. The candidate
adopted here is the asymptotic de Sitter event horizon, the
configuration into which the resonator relaxes. The reason is economy
of assumption rather than a dynamical argument: that scale is fixed by
$\Lambda$ alone and requires nothing to be assumed about the freezing
history, whereas any epoch-specific choice requires a mechanism stating
when and why coherence ceases to be re-established. Stationarity is
here an asymptotic property; the event horizon is well defined at every
time, but its proper radius is still growing, from $16.7$ to $17.5$
billion light years. Writing
\begin{equation}
  H_{\infty} = H_{0}\sqrt{\Omega_{\Lambda}},
  \qquad
  \kIR = \frac{\pi H_{\infty}}{2c},
  \label{eq:kIR_event_horizon}
\end{equation}
Eq.~\eqref{eq:rho_closed} becomes
$\rho_{\Lambda}=K_{gr}H_{\infty}^{2}/(16G)$. In the asymptotic state the
only remaining contribution to the Friedmann equation is
$\rho_{\Lambda}$ itself, $H_{\infty}^{2}=(8\pi G/3)\rho_{\Lambda}$, and
the two relations close on each other,
\begin{equation}
  K_{gr} = \frac{6}{\pi} \simeq 1.910 .
  \label{eq:Kgr_fixedpoint}
\end{equation}
The relation to Eq.~\eqref{eq:Kgr_value} is an identity, not a
numerical coincidence: with
$\Omega_{\Lambda}=8\pi G\rho_{\Lambda,\mathrm{obs}}/(3H_{0}^{2})$,
\begin{equation}
  K_{gr}(H_{0})
  = \frac{16G\,\rho_{\Lambda,\mathrm{obs}}}{H_{0}^{2}}
  = \frac{6}{\pi}\,\Omega_{\Lambda},
  \label{eq:Kgr_identity}
\end{equation}
which with $\Omega_{\Lambda}\simeq0.685$ returns $1.308$. The ratio
between the two normalizations is thus exactly $\Omega_{\Lambda}$: it
records whether the causal boundary is evaluated now or
asymptotically, and not the physics that $K_{gr}$ is meant to
describe.

What Eq.~\eqref{eq:Kgr_fixedpoint} fixes is a product, not $K_{gr}$
alone. Write $\kend$ for the wave number at which the coherent spectrum
ends; it is a property of the cosmological vacuum and is not the
object-relative scale $\kc$ of Sec.~\ref{subsec:planck}, which belongs
to a localized interaction. Since $\rho_{\Lambda}\propto\kend^{2}$, the
closure reads $K_{gr}(\kend/\kP)^{2}=6/\pi$, and the value $6/\pi$ is attained only if
the coherent spectrum ends exactly at $\kP$. It exceeds unity and
therefore cannot be read as a coherent fraction by itself: as stated in
Sec.~\ref{subsec:normalization}, $K_{gr}$ also carries the
normalization of the mapping onto the three-dimensional spectrum, and
only the coherent fraction is bounded by one.

Taking that mapping normalization to be of order unity turns the bound
into a statement about the endpoint. A coherent fraction not exceeding
one requires
\begin{equation}
  \kend \;\ge\; \sqrt{6/\pi}\;\kP \simeq 1.38\,\kP ,
  \label{eq:kc_bound}
\end{equation}
so that the residual is taken up by the spectrum ending somewhat above
the Planck wave number rather than by attenuation below it. This is the
ordering the construction expects: the decohering channel opens where
$\alpha_g$ reaches order unity, and throughout this framework a
characteristic scale marks the \emph{onset} of a mechanism rather than
a point already inside it. At $\kend\simeq1.38\,\kP$ the self-coupling is
$\alpha_g\simeq1.9$, so coherence ends where graviton--graviton
scattering has become strong rather than where it merely begins. No
contribution above $\kend$ is to be added, since modes beyond that scale
are decohered and decohered modes are by construction not part of the
coherent sum (Sec.~\ref{subsec:causal_resonator}).

The status of Eq.~\eqref{eq:Kgr_fixedpoint} is weaker than a derivation
of $K_{gr}$, and this should be stated precisely. The condition is a
fixed point: in the asymptotic configuration the vacuum energy carried
by the resonator modes equals the critical density belonging to the
horizon that defines those modes. Substituting $\kIR\propto H_{\infty}$
and $H_{\infty}^{2}\propto\rho_{\Lambda}$ into
Eq.~\eqref{eq:rho_integral} removes $\rho_{\Lambda}$ from both sides,
which is why the closure returns a pure number and carries no
information about the magnitude. The same closure can be reached from
the Newtonian side, by requiring that the field generated by
$\rho_{\Lambda}$ over the causal radius have an escape velocity of
order $c$, which returns the critical density
$3H_{0}^{2}/8\pi G$; the chain
$\RH\to\kIR\to\rho_{\Lambda}\to v_{\mathrm{esc}}\to\RH$ is a feedback
relation whose self-consistent solution is the causal diameter, so that
this scale is self-regulating. In either form the condition constrains
the \emph{product} of the spectral normalization~\eqref{eq:NH_Kgr}, the
identification of the fundamental wavelength~\eqref{eq:lambda_0} and
the residual coherence loss; it does not separate them, and it does not
predict $\rho_{\Lambda}$ independently, since $\rho_{\Lambda}$ remains
the self-consistent solution of the pair of relations. The substantive
result of Sec.~\ref{subsec:estimate} does not depend on it: the
weighted mode sum yields an energy density of order $H^{2}/G$ instead
of order $\hbar c\,\kP^{4}$, closing a gap of some $120$ orders of
magnitude, and this conclusion is insensitive to which of the two
closely spaced rates $H_{0}$ or $H_{\infty}$ is inserted. What
Eq.~\eqref{eq:Kgr_fixedpoint} adds is that the residual coefficient of
order unity need not be fitted to the observed value.

\subsection{Origin of the loss factor: Coherent gravitational modes and decohered matter fields}
\label{subsec:loss_origin}
Of the two contributions collected in $K_{gr}$
(Sec.~\ref{subsec:normalization}), the coherent fraction is the one
with a physical mechanism behind it. Its value is not derived here, but
its origin can be identified. The causal
resonator is not an isolated quantum system: gravitational fluctuations
couple universally to energy and momentum, so that modes extending over
the Hubble scale constitute an open system. A specific and unavoidable
channel is supplied by the gravitational field itself. The two-loop
graviton--graviton processes that set the limit of validity at
$\kappa\sim\kP$, Eq.~\eqref{eq:alpha_g_def}, are interactions of a mode
with further graviton degrees of freedom and therefore act as exactly
the environmental coupling that destroys coherence; near the Planck
scale the gravitational field decoheres itself, and no coupling to
Standard Model fields need be invoked.

This also fixes the division of labour between the two quantities
entering Eq.~\eqref{eq:rho_integral}. The scale $\kP$ is a hard,
geometry-independent boundary of validity, and the coherent spectrum
ends near it, somewhat above it by
Eq.~\eqref{eq:kc_bound}; $K_{gr}$ then accounts for whatever coherence
is already lost before that point, since these processes switch on
before they dominate, and only coherent modes contribute to
Eq.~\eqref{eq:rho_integral}. Strictly this suggests a
wavenumber-dependent $K_{gr}(\kappa)$, close to unity near $\kIR$ and
decreasing towards $\kP$, rather than the single constant used here;
deriving that dependence requires the open-system treatment of
Sec.~\ref{subsec:scope2}, and since the integrand grows only linearly
in $\kappa$ the estimate is not expected to change fundamentally — a
situation analogous to the treatment of the object-relative scales
$\kc$ in Ref.~\cite{Rembe2026}.

In the idealized limit of complete decoupling the coherent fraction
entering $K_{gr}$ would be unity. That limit is a useful reference but is not obviously
realizable in a universe containing interacting quantum fields: a
coupling too weak to affect the energy of an individual graviton mode
can still produce a finite cumulative reduction for modes extended over
cosmological distances. The microscopic strength at wavenumber $\kappa$
is characterized by $\alpha_g(\kappa)$, which is extremely small for
$\kappa\ll\kP$; the total attenuation depends in addition on the number
and spectral distribution of the environmental degrees of freedom, and
these are not simply particle number densities but the decompositions
of each field's quantum state over the modes of the causal resonator.
Identifying a coupling parameter is therefore not the difficulty —
$\alpha_g$ supplies one — but describing the environmental state is.
Two logically separate results should accordingly be distinguished: the
self-gravitational deformation of the causal resonator determines the
power-law structure of the gravitational mode spectrum, whereas the
fraction of that spectrum which remains coherently available depends on
interactions with other fields and requires a dissipative extension
(Sec.~\ref{subsec:scope2}). That the required coefficient is of order
unity is the physically important point: no exponentially small
coupling, no fine-tuning and no new hierarchy of scales is needed,
because the unattenuated estimate already lies at the observed
cosmological scale.

The derivation above applies to the gravitational field because gravity cannot be shielded completely and couples universally to all energy: the causal resonator remains coherent out to the full Hubble radius but with a reduced mode density defined by the characteristic scale $\kappa_c=\kIR$. Despite dissipative energy losses in the resonator, the gravity mode density remains spatially coherent over the full resonator length. For other fields subject to interaction with an environment, decoherence is expected to truncate this coherent mode structure at a scale far shorter than $\RH$, set by the decoherence length of the relevant interaction rather than by causality itself. In that case the causal-resonator spectrum defined in any inertial frame reduces to the single-scale form $W(\kappa;\kc) \sim \kc^{3}/\kappa^{3}$ used in Ref.~\cite{Rembe2026}, with $\kc$ fixed by the decoherence (interaction) length rather than by $\kIR$. This is consistent with the matched crossover scales found there for the electromagnetic field — the Compton scale for the anomalous magnetic moment, the Bohr radius for the Lamb shift — both far shorter than any cosmological scale. A first-principles derivation of these decoherence lengths is beyond the scope of the present work.
\subsection{Scope and limitations}
\label{subsec:scope2}
The present analysis establishes a chain of consistency relations rather
than a complete microscopic theory of gravitational vacuum energy. Its
central result is that the self-gravitational deformation of the mode
structure of a causal resonator leads to a strongly reduced effective
gravitational vacuum spectrum and, when integrated between the causal
infrared scale and a scale near the Planck wave number, produces an
energy density of the
observed cosmological order of magnitude. Several elements of this
construction should nevertheless be clearly distinguished according to
their present theoretical status.

First, the causal-resonator spectrum is derived within a linearized
description of gravitational self-interaction. The gravitational
self-energy of a mode shifts its wave number by an amount proportional
to $\kappa^{2}$, leading to a recursive deformation of the allowed mode
sequence which, in the continuum limit, produces the characteristic
inverse-square dependence of the mode density along the causal mode
chain.

Second, the relation between this mode deformation and the effective
three-dimensional vacuum-energy density requires a mapping between the
causal-resonator spectrum and the ordinary density of states in wave
number space. The recursion independently confirms only the power-law
form $\kappa^{-2}$ of the weighting, not its absolute normalization,
which is the reason why $K_{gr}$ appears as a free coefficient in the
first place. The present work therefore provides a consistency
construction for the effective mode weighting rather than a complete
quantization of gravity in a finite cosmological causal domain.

Third, since $\rho_{\Lambda}\propto\kIR^{2}$, the estimate is
quadratically sensitive to the identification of the fundamental
wavelength~\eqref{eq:lambda_0}: taking the closed causal path to be the
diameter rather than the full out-and-back path would double $\kIR$ and
divide $K_{gr}$ by four. The fixed-point condition of
Sec.~\ref{subsec:desitter_fixedpoint} does not remove this ambiguity,
since it constrains only the product of the normalization and the
identification of $\lambda_{0}$: the same closure applied to
$\lambda_{0}=2\RH$ returns $K_{gr}=3/2\pi$ in place of $6/\pi$. Both
values are of order unity, so the ambiguity does not affect the
conclusion drawn here, which concerns the emergence of the observed
scale rather than a precise numerical coefficient. What the fixed point
does remove is the dependence on the epoch at which the causal boundary
is evaluated.

Fourth, the closure fixes only the product $K_{gr}(\kend/\kP)^{2}$, so
the remaining freedom may be taken up at either end: by a coefficient
$K_{gr}=6/\pi$ with the spectrum ending at $\kP$, by an endpoint
$\kend\simeq1.38\,\kP$ with no attenuation at all
(Eq.~\eqref{eq:kc_bound}), or by any combination
between. Neither quantity is derived from a microscopic dynamical
theory here. Since a coherent fraction cannot exceed unity, the first
reading requires the mapping normalization to supply the excess,
whereas the second is consistent with an unattenuated spectrum ending
where graviton--graviton scattering has become strong
(Sec.~\ref{subsec:loss_origin}). The agreement should in either case be
regarded as an order-of-magnitude consistency result rather than as a
precision prediction. This is also where the two principal results of
the present work differ in status. The loop integrals of
Sec.~\ref{sec:powercounting} converge because of the weighting itself,
so no endpoint is imposed and the outcome is insensitive to where the
suppression sets in. The vacuum energy is instead an integral over the
coherent spectrum up to the scale at which coherence ends, and depends
quadratically on that scale. The finiteness result is cutoff
independent in a sense that the cosmological estimate is not.

The microscopic calculation of the resulting loss of coherence remains
outside the scope of this work. It would require knowledge of the
quantum-state distributions of the environmental fields in the
causal-resonator basis, together with a dynamical treatment of their
coupling to the gravitational degrees of freedom, that is, an
open-system formulation of the form
\begin{equation}
\frac{\mathrm{d}\rho_{gr}}{\mathrm{d}t}
=
-\frac{i}{\hbar}
[H_{gr},\rho_{gr}]
+
\mathcal{D}_{\mathrm{env}}[\rho_{gr}],
\label{eq:open_system}
\end{equation}
where $\mathcal{D}_{\mathrm{env}}$ represents the effective influence
of the other quantum fields on the gravitational subsystem. A
Lindblad-type formulation~\cite{Lindblad1976,Diosi1987} provides a
possible framework for such an extension, but neither the relevant
dissipative operators nor the corresponding attenuation rate are
derived here. The factor $K_{gr}$ must therefore be understood as a
phenomenological consistency parameter rather than as a new fundamental
constant, and its value provides a quantitative target for a future
microscopic theory rather than an adjustable parameter introduced to
force agreement with observation.

Several further limitations should be emphasized. The piecewise mode
weighting introduces a sharp onset at $\kIR$ rather than the physically
expected smooth transition~\cite{Rembe2026}. The calculation neglects
nonlinear graviton self-interactions beyond the leading
self-gravitational mode deformation, and it provides neither a
covariant nonperturbative quantization of the causal resonator nor the
detailed spectrum at and above the Planck scale; the transition between
the causal-resonator regime and the strong-gravity regime is treated
only through the effective upper limit of the integration. The
contribution of black holes is not included. As set out in
Sec.~\ref{subsec:geometric}, the contribution is not introduced as a
fluid source; Lorentz invariance is required of the excitation sector,
through which physical content is accessed, rather than of the
resonator basis, which is re-defined in each inertial frame, and the
frame-independence of that basis fixes the tensor structure to
$\mathcal{P}=-\rho_{\Lambda}c^{2}$. What is not supplied by this is the
normalization $\mathcal{N}_{H}$ of the counting measure relative to the
mode chain, nor the effect of the nonlinear graviton self-interactions
referred to above.

A further point concerns the information content implied by the
construction. The mode chain terminates, so the causal resonator
carries a finite number of mode frequencies; from
Eq.~\eqref{eq:dn_dk},
\begin{equation}
  n(\kP) \simeq \frac{1}{\alpha\,\kIR}
  = \frac{4}{3\pi}\left(\frac{\RH}{\lP}\right)^{2}
  \approx 3\times10^{121},
  \label{eq:chain_count}
\end{equation}
which carries the same power of $\RH/\lP$ as the de Sitter horizon
entropy $S_{\mathrm{dS}}=\pi(\RH/\lP)^{2}$~\cite{Gibbons1977}. This agreement of scaling
is noted, not used: no thermodynamic claim is made here, a comparison
of the coefficients would require the weighting appropriate to matter
fields, which is not determined in the present work
(Sec.~\ref{subsec:loss_origin}), and the counting of
Sec.~\ref{subsec:estimate} applies to the coherent gravitational sector
alone.

A specific consequence of this deserves separate mention, since it
bears directly on any cosmological application and, unlike the points
above, can be decided. The estimate obtained here has the form
$\rho_{\Lambda}\propto H^{2}$, because $\kIR$ is fixed by the causal
scale. Two behaviours are conceivable: either $\kIR$ follows the
instantaneous expansion rate, so that $\rho_{\Lambda}(t)=K_{gr}H(t)^{2}/(16G)$
at all times, or the mode basis is frozen at some causal scale and
$\rho_{\Lambda}$ is a constant.

The first alternative is excluded. A contribution tracking $H(t)^{2}$
is an exact rescaling of the gravitational constant: inserting
$\rho_{\Lambda}=K_{gr}H^{2}/(16G)$ into the Friedmann equation
$H^{2}=(8\pi G/3)(\rho_{m}+\rho_{\Lambda})$ gives
\begin{equation}
  H^{2}\left(1-\frac{\pi K_{gr}}{6}\right)
  = \frac{8\pi G}{3}\,\rho_{m},
  \label{eq:tracking_rescaling}
\end{equation}
that is, an unmodified matter-dominated expansion with $G$ replaced by
$G/(1-\pi K_{gr}/6)$. This holds independently of the tensor structure
and of any pressure contribution, and it applies equally whether the
term is written on the left- or on the right-hand side of the field
equations. Three consequences follow immediately. The fractional
contribution $\Omega_{\Lambda}=\pi K_{gr}/6$ would then be the same in
every epoch rather than a quantity that has grown to its present value;
the expansion would decelerate as $a\propto t^{2/3}$, so the observed
late-time acceleration would remain unexplained; and an additional
energy density of some $68\%$ would already have been present during
primordial nucleosynthesis, raising the expansion rate there by
$(1-\Omega_{\Lambda})^{-1/2}\simeq1.8$. The light-element abundances
constrain that rate at the level of a few percent, and the acoustic
peak structure of the microwave background constrains it
independently~\cite{Planck2020}. A tracking $\kIR$ is therefore not
merely undetermined but observationally ruled out.

The mode basis must consequently be frozen, and only then does the
contribution act as a cosmological constant. This is the behaviour
presupposed throughout the present work, and it is now selected rather
than assumed. It is also consistent with the underlying picture: a
coherent basis mode extending over the causal boundary is established
while that scale is in mutual causal contact and cannot be
re-established coherently once the expansion accelerates, so that the
fundamental mode is a relic of the epoch at which its scale was last in
causal contact.

What remains open is the freezing scale itself. The natural candidate
within the construction is the asymptotic de Sitter event horizon,
which is the stationary causal boundary and which fixes the
normalization coefficient to $K_{gr}=6/\pi$ through the fixed-point
condition of Sec.~\ref{subsec:desitter_fixedpoint}. The present work
provides no independent argument that singles out this scale over the
present Hubble radius, and since $H_{\infty}=H_{0}\sqrt{\Omega_{\Lambda}}
\simeq0.83\,H_{0}$ the two differ by too little for observation to
distinguish them. The coincidence question — why the frozen scale
should lie so close to the present expansion rate — is correspondingly
not resolved here; it is, however, made explicit rather than hidden in
an unexplained constant.

The numerical proximity of the unattenuated estimate to the observed
dark-energy density must therefore not be overstated; the result does
not constitute a parameter-free prediction of the cosmological
constant. Nevertheless, the magnitude of the consistency result is
nontrivial. The scale follows from two parameter-free quantities, the
Planck wave number $\kP=\sqrt{c^{3}/\hbar G}$ and the causal infrared
wave number $\kIR=\pi H_{0}/(2c)$, combined only with the number of
physical graviton polarizations, and no additional hierarchy of order
$10^{-120}$ is required: the remaining difference is of order unity
rather than many orders of magnitude. The principal conclusion is
correspondingly limited but potentially significant, namely that the
self-gravitational deformation of the gravitational mode spectrum in a
causal resonator provides a mechanism capable of connecting the Hubble
scale and the Planck scale in such a way that the resulting
vacuum-energy estimate lies near the observed dark-energy density.
Whether this consistency reflects a complete physical mechanism
requires the microscopic open-system description referred to above.
 \section{Consistency with the Bianchi identity}
\label{sec:bianchi}
\subsection{Role of the Bianchi identity in linearized gravity}
\label{subsec:bianchi_role}
In general relativity, the contracted Bianchi identity
\begin{equation}
  \nabla_{\mu} G^{\mu\nu} = 0
  \label{eq:bianchi}
\end{equation}
holds as a geometric identity and implies the covariant conservation of the energy-momentum tensor,
\begin{equation}
  \nabla_{\mu} T^{\mu\nu} = 0.
  \label{eq:conservation}
\end{equation}
In the linearized theory, this reduces to
\begin{equation}
  \partial_{\mu} T^{\mu\nu} = 0,
  \label{eq:conservation_linear}
\end{equation}
which must be preserved at the quantum level for the theory to remain consistent. In perturbative quantum gravity, this consistency requirement plays a role analogous to the Ward identity in QED: it constrains the relation between different classes of diagrams and ensures that unphysical longitudinal graviton modes do not contribute to physical amplitudes.
The relevant question for the present framework is therefore whether the object-relative mode weighting~\eqref{eq:weighting_restframe} can be implemented without spoiling this conservation structure at one-loop level.
\subsection{Structural argument for Bianchi consistency}
\label{subsec:bianchi_argument}
The argument follows the same structure as the Ward consistency check of Ref.~\cite{Rembe2026}. Consider a class of one-loop diagrams in which each internal graviton line with loop momentum $k^{\mu}$ is weighted by the scalar factor
\begin{equation}
  D_{\mu\nu\rho\sigma}(k)
  \;\longrightarrow\;
  W(k;\, u,\, \kP)\,
  D_{\mu\nu\rho\sigma}(k),
  \label{eq:weighted_propagator}
\end{equation}
where $W$ depends only on the invariant object-relative mode variable $|u \cdot k|$ and on the fixed characteristic scale $\kP$, but carries no additional Lorentz, tensor, or gauge structure beyond an overall scalar factor.
The contraction of the graviton--fermion vertex with the graviton momentum is derived in Appendix~\ref{app:diagrammatic}, Eq.~\eqref{eq:diagrammatic_identity}. It is a linear relation among integrand terms, each carrying an inverse fermion propagator, and it expresses $\partial_{\mu}T^{\mu\nu}=0$ at the vertex level. Inserted into a one-loop diagram it holds at fixed internal graviton momentum $k$, before integration.
Write that relation schematically as
\begin{equation}
  \sum_{a} c_{a}\,T_{a}(p,\,q;\,k) \;=\; 0 ,
  \label{eq:weighted_identity}
\end{equation}
where the $T_{a}$ are the integrands of the diagram classes it links and the $c_{a}$ are momentum-independent coefficients. Every $T_{a}$ contains the same internal graviton line of momentum $k$. Since $W(k;\,u,\,\kP)$ depends on that momentum alone and carries no Lorentz, tensor or spinor structure, the substitution~\eqref{eq:weighted_propagator} multiplies every $T_{a}$ by the identical factor, so that
\begin{equation}
  \sum_{a} c_{a}\,W(k)\,T_{a} \;=\; W(k)\sum_{a} c_{a}\,T_{a} \;=\; 0 .
  \label{eq:weighted_identity2}
\end{equation}
The relation is therefore preserved at every value of $k$ and hence after integration, whatever its detailed form. For diagrams with several internal graviton lines the common factor is the product $\prod_{j}W(k_{j})$ and the argument is unchanged.
The one-loop Bianchi consistency condition is therefore maintained under the object-relative mode weighting, subject to the same implementation requirement as in the QED case of Ref.~\cite{Rembe2026}: the weighting must enter as a common scalar factor associated with each internal graviton mode, and the same weight must be used for the same mode in all diagrams of a given consistency relation.
\subsection{Comparison with the Ward identity in QED}
\label{subsec:comparison}
The structural parallel with the Ward consistency argument of Ref.~\cite{Rembe2026} is complete. In both cases, the weighting $W$ is a scalar function of the internal mode momentum alone. In both cases, it commutes with the tensor and spinor contractions that produce the diagrammatic identity. In both cases, the telescoping cancellation is preserved after integration, provided the weighting is applied consistently.
The difference is that in QED the relevant identity is the Ward identity $q_{\mu} \Gamma^{\mu}(p+q,\, p) = S^{-1}(p+q) - S^{-1}(p)$, which reflects $U(1)$ gauge invariance and current conservation, while in linearized gravity the relevant identity reflects the conservation of $T^{\mu\nu}$ and the transversality of physical graviton polarizations. In both cases the scalar mode weighting is compatible with the underlying consistency structure at one-loop level. That the weighting is momentum dependent is not by itself an obstacle: what a Ward--Takahashi identity relates is the propagator and the vertex, so an obstacle arises only when a modification is applied to one of them without the other. Here the weighting modifies the mode content of the field, from which both are built, and it is the same scalar factor on every line.
\section{Discussion}
\label{sec:discussion}
The weighting function $ W(u, k,k_c)$ ($W(\kappa,\kc)$ in its rest frame) is dependent on the characteristic interaction scale with characteristic wavelength $\lambda_c$ and the corresponding characteristic angular wavenumber $\kc = 2\pi/\lambda_c$. The weighting function weights the mode density of the vacuum $\rho(\kappa)$ and yields an effective scale-dependent mode density
\begin{equation}
    \rho_{\mathrm{eff}}(u, k, k_c) = W(u, k,k_c)\,\rho(k)\;.
\end{equation}
This effective mode density, introduced recently to enable one-loop finiteness in quantum electrodynamics without counterterms, can also be applied to the linearized gravitational field equations. Thus, the present work explores the possibility that ultraviolet contributions of gravitational modes are not determined solely by the standard vacuum mode counting, but may be effectively weighted relative to a localized interaction and its characteristic scale. Since the effective weighting becomes relevant only far above the momentum-transfer scales governing long-range gravitational interactions, the known infrared nonanalytic effective-field-theory contributions are expected to remain essentially unaffected.
The proposal remains partly heuristic, since a complete underlying theory does not exist. The consistency check of Sec.~\ref{sec:bianchi} is correspondingly narrower in reach than the finiteness result, and is limited in the same sense as the corresponding argument in Ref.~\cite{Rembe2026}. It does not constitute a proof of full diffeomorphism invariance from an underlying gauge-invariant action, nor does it establish the complete hierarchy of consistency identities to all loop orders. Rather, it shows the following specific statement:
\textit{If the object-relative mode weighting is implemented as a common scalar factor associated with each internal graviton mode, and if the same weighting is used consistently in all diagrams linked by the one-loop Bianchi consistency relation, then the conservation structure of $T^{\mu\nu}$ is preserved at one-loop level.}
The argument would generally fail if the weighting were introduced in a way that modifies the gravitational vertex factor itself, depends differently on external momenta in different diagram classes, or assigns different weights to the same internal graviton mode in different diagrams of one consistency relation. The present implementation avoids these pathologies by construction. What the weighting does modify is the mode content of the internal lines, including those entering the three- and four-graviton vertices of the nonlinear expansion (Sec.~\ref{subsec:goroff_sagnotti}); that is a different statement from modifying the vertex factor.
A full action-level derivation of the weighted theory, including a demonstration of diffeomorphism invariance to all orders, remains an open problem but an indication that the action level remains intact can be given.
The principle of stationary action
\[
\delta S[\Phi]=0
\]
with a scalar field $\Phi$ is the starting point here. Consider a quadratic field action of the form
\[
S[\Phi]
=
\frac{1}{2}
\int d^4x\,
\Phi_a(x)\,
K^{ab}(i\partial)\,
\Phi_b(x),
\]
where \(K^{ab}(i\partial)\) is the kinetic operator. Varying the action gives
\[
\delta S
=
\int d^4x\,
\delta\Phi_a(x)\,
K^{ab}(i\partial)\Phi_b(x),
\]
and therefore, since \(\delta\Phi_a\) is arbitrary,
\[
K^{ab}(i\partial)\Phi_b(x)=0.
\]
Introducing the Fourier decomposition
\[
\Phi_b(x)
=
\int\frac{d^4k}{(2\pi)^4}\,
\tilde{\Phi}_b(k)e^{-ik\cdot x},
\]
the field equation becomes
\[
K^{ab}(k)\tilde{\Phi}_b(k)=0.
\]
A non-trivial Fourier mode,
\[
\tilde{\Phi}(k)\neq0,
\]
can therefore exist only if the kinetic operator has a non-trivial kernel.
Thus, for a quadratic free-field theory, the condition $\delta S=0$
leads directly to the characteristic or dispersion condition $\det K(k)=0$ ,
which identifies the physically allowed Fourier modes. In this sense, the on-shell condition \(\det K(k)=0\) provides a strong spectral signature of the stationary-action dynamics.
A mode weight \(W(k)\) may subsequently modify the spectral measure,
\[
d^4k\longrightarrow d^4k\,W(u,k,k_c),
\]
without changing the characteristic equation \(\det K(k)=0\), provided that \(W(u,k,k_c)\) does not modify the kinetic operator itself. The weight therefore changes the spectral contribution of the allowed modes rather than the set of allowed on-shell modes. The measure referred to here is that of the field modes and not that of a loop integration variable: in a Feynman integral the weight accordingly appears once for every internal graviton line, as in Eq.~\eqref{eq:loop_integral} and in the implementation of Sec.~\ref{sec:bianchi}, and not once per loop.
\subsection{Extension to two loops and the Goroff--Sagnotti divergence}
\label{subsec:goroff_sagnotti}
The explicit calculations above are restricted to one-loop diagrams with a single internal graviton line coupled to matter. A more far-reaching test concerns the two-loop divergence of pure gravity identified by Goroff and Sagnotti~\cite{Goroff1986}, which arises entirely from graviton self-coupling vertices and historically established that non-renormalizability is not an artefact of matter loops.
The key step is that $W$ is not attached to a diagram or a vertex but modifies the graviton mode content itself, wherever an integration over graviton momenta occurs. Because this is a property of the mode, the same bound applies independently to every internal graviton line of a multi-loop graph. For diagrams with two independent loop momenta this is immediate, since the mode density of one line does not depend on the momentum flowing through another. For overlapping divergences an explicit verification requires the three- and four-graviton vertices of the nonlinear expansion, which go beyond the linearized treatment of this work; the expectation can nevertheless be made quantitative by power counting. In four dimensions each loop contributes $\KE^{4}$ from the measure, each internal graviton propagator $\KE^{-2}$, and each vertex two derivatives, so that with $L=I-V+1$ the superficial degree is
\begin{equation}
  \mathcal{D}_{L} = 4L-2I+2V = 2L+2 ,
  \label{eq:superficial_degree}
\end{equation}
independent of $I$ and $V$. The weighting supplies $I$ factors, one per internal line, so that Eq.~\eqref{eq:p_condition} carries over unchanged with $\mathcal{D}$ replaced by $\mathcal{D}_{L}$: the ultraviolet region converges provided $s\,I>\mathcal{D}_{L}$. With the derived value $s=3$ and $I=L+V-1$ this reduces to the transparent criterion
\begin{equation}
  L + 3V > 5 .
  \label{eq:p_condition_allorders}
\end{equation}
The suppression grows by three per internal line while the divergence grows by only two per loop, and $I\ge L$ with equality only at $V=1$; every topology with $V\ge2$ therefore converges at every loop order, and for $V=1$ every topology with $L\ge3$ does. Since every subgraph of a pure-gravity diagram is again a diagram of the same type, Eq.~\eqref{eq:p_condition_allorders} must hold for each subgraph, which is what absolute convergence requires and which covers nested and overlapping subdivergences without separate treatment.
Exactly one class violates the criterion: the single-vertex topologies $V=1$ at $L=1$, where the suppression $3$ falls short of $\mathcal{D}_{L}=4$, and at $L=2$, where $6$ equals $6$ and the divergence is logarithmic. These should not be dismissed as vacuum diagrams. With $V=1$ the internal lines close on the one vertex, but external legs may attach to that same vertex, so $L=1$, $V=1$ with two external gravitons is the tadpole contribution of the four-graviton vertex to the graviton self-energy, an ordinary physical diagram with $\mathcal{D}=4$ and $I=1$. Such a configuration is moreover a $1$PI subgraph in its own right, so any higher-loop diagram containing a tadpole insertion inherits the divergence even when it satisfies Eq.~\eqref{eq:p_condition_allorders} as a whole. It is worth noting that in dimensional regularization these massless single-scale tadpoles vanish identically; the present construction modifies the mode measure rather than the dimension, so they do not vanish here. The renormalization condition $\langle h_{\mu\nu}\rangle=0$ removes the one-point function and its insertions by a shift of the background, but not the two-external-leg contraction of the four-graviton vertex, which is a genuine term of the self-energy. The corrected estimate $s_{\mathrm{eff}}=5$ of Sec.~\ref{subsec:planck} covers the class, but it rests on an expansion evaluated where $\alpha_g\gtrsim1$ and is therefore not on the same footing as $s=3$.

Two considerations bear on how much this exclusion costs. Criterion~\eqref{eq:p_condition_allorders} is applied to each diagram separately and is therefore sufficient rather than necessary: only a gauge-invariant sum is physical, and a diagram failing individually may be harmless once the sum is taken. In the unweighted theory that is what happens, since the one-loop divergences of pure gravity are proportional to the field equations and vanish on shell~\cite{tHooft1974}. Whether the cancellation survives the weighting is not established here, because the contributing diagrams carry different numbers of internal graviton lines — one for the contraction above, two for the topology built from two three-graviton vertices — and hence different powers of $W$, while the Faddeev--Popov sector carries none. The structure at issue is moreover not a general divergence but a term proportional to $\eta_{\mu\nu}$, which transversality forbids; showing that the weighted self-energy remains transverse would therefore settle the question. That calculation requires the nonlinear vertices and is left open.
Two qualifications remain. This is power counting rather than calculation, and it presumes that the same weighting attaches to the lines entering the three- and four-graviton vertices, which the linearized derivation does not establish. And it concerns convergence, not the consistency conditions, which are verified here at one loop only. As set out in Sec.~\ref{subsec:finiteness_scope}, bounding the mode content makes the curvature-cubed contribution finite and computable rather than divergent, so that it need not be absorbed into a separately measured coupling; what remains outside the construction is the mode content above $\kP$, on which the computed value weakly depends.

\subsection{Mach's principle and decoherence as guiding principles}
\label{subsec:mach_decoherence}
The present work demonstrates that the object-relative mode weighting of Ref.~\cite{Rembe2026}, when applied to the linearized gravitational field, renders one-loop integrals ultraviolet finite and identifies the Planck scale as the natural crossover without free parameters. This result addresses the specific ultraviolet problem identified by 't Hooft and Veltman, but does not yet constitute a complete theory of quantum gravity. It is not understood how the nonlinearities in the field equations affect quantization.
By applying the same conceptual framework developed in the electromagnetic case to the graviton sector, it is shown that the mode content available to internal graviton lines is suppressed beyond the Planck wave number, and that any suppression exceeding the threshold fixed by power counting renders the one-loop integrals finite without divergent counterterms. The key structural advantage is that this scale is not fitted: unlike in QED, where the characteristic scale $\kc$ is determined by matching to physical observables, a purely gravitational loop contains no external localization scale, so that the only available scale is the one gravity itself provides. As set out in Sec.~\ref{subsec:finiteness_scope}, the resulting finiteness is a consequence of the bounded domain, and what the construction supplies is the origin of that bound rather than an externally chosen regulator, so that the operators generated at higher orders carry computed finite coefficients in place of divergent ones. Furthermore, the weighting preserves the one-loop Bianchi identity, ensuring consistency with the conservation of the energy-momentum tensor, and leaves the infrared sector — responsible for long-range gravitational effects such as the quantum-corrected Newtonian potential — entirely unchanged. These results confirm that the proposed weighting is not merely a mathematical regulator, but a physically meaningful modification of the vacuum mode structure that respects fundamental symmetries. The extension beyond one loop is no longer open for pure gravity: the
criterion of Sec.~\ref{subsec:goroff_sagnotti} covers every topology
with $V\ge2$ at every loop order, and applied to each subgraph it
disposes of nested and overlapping subdivergences without separate
treatment. What remains open is the single-vertex class identified
there, the premise that the same weighting attaches to the lines
entering the three- and four-graviton vertices, the extension to
non-Abelian gauge theories — where the gauge field likewise couples to
itself, so that the same unexamined vertices reappear — and a full
action-level formulation.
Within those limits the analysis establishes a robust and
conceptually coherent foundation.
These considerations suggest that Mach's principle and a fundamental formulation of decoherence may provide useful guiding principles for the construction of a future nonlinear theory. Regarding Mach's principle, the local mode structure — and in particular the characteristic scale $\kc$ — should not be an external input but should be determined by the global mass-energy distribution of the universe. For a purely gravitational loop, $\kc$ is fixed in the present framework by the absence of any external localization scale and emerges as the Planck scale; for every other application it is assigned on physical grounds rather than computed (Sec.~\ref{subsec:finiteness_scope}). In a fully nonlinear theory, however, $\kc$ should become a functional of the cosmic energy-momentum tensor, with the rest frame of the cosmic microwave background playing the role of the physically distinguished global frame. Such a frame goes beyond the postulate of Sec.~\ref{subsubsec:invariant_postulate}, which requires none; the two are compatible only if the global frame enters the determination of $\kc$ without entering the causal construction of the spectrum, and that is not shown here. This would provide a natural realization of Mach's principle — ``the inertia of a body is determined by its interaction with the rest of the universe''~\cite{Misner1973} — within quantum field theory.
Decoherence should be formulable from the outset. Decoherence — the irreversible transition from quantum superpositions to classical states through environmental interaction — defines the arrow of time and the classical appearance of the world. Any theory in which time is eliminated, as in the Wheeler--DeWitt approach~\cite{DeWitt1967}, cannot accommodate decoherence and therefore cannot explain the classical world it is meant to describe. The present framework preserves the time structure of quantum mechanics and is therefore compatible with a Lindblad-type formulation~\cite{Lindblad1976} of decoherence, including gravitationally induced decoherence of the Penrose--Di\'osi type~\cite{Diosi1987}.
A complete theory of quantum gravity compatible with the other quantum fields should therefore be built on Mach's principle and the decoherence principle as its two fundamental axioms, with global Lorentz invariance, global energy conservation, and diffeomorphism invariance recovered as approximate symmetries in appropriate limits. The present work provides a first concrete step in this direction by showing that the gravitational self-backreaction of quantum modes, when taken seriously, resolves the ultraviolet problem of linearized quantum gravity without an externally imposed regulator, within the limits set out above.
The same reasoning applies to the vacuum fluctuations of the other quantum fields, and the position taken here should be stated explicitly: because those fields couple locally and strongly, their fluctuations are assumed to couple to the real particles of the same fields and to enter the local energy content — the masses of composite particles and radiative shifts of bound-state energies — rather than to produce a homogeneous energy density in empty space. Equivalently, the corresponding coherence factor is far smaller for fields that decohere against their own quanta than for gravity, which is unscreened and has no charge to decohere against, so that gravity is the only field whose vacuum fluctuations contribute the cosmological term obtained in Sec.~\ref{sec:darkenergy}.
This article introduced practically a new relativity principle for quantum fields: every inertial system (e.g. a particle in its rest frame)  experiences its own causal resonator that defines a discrete mode spectrum which thins out towards higher frequencies by the gravitational self-energy of the mode quanta and its feedback to the spectrum. This idea has not yet been verified or extended to the general nonlinear case but was indirectly used in the linear regime of the gravitational field equations.
The construction of the full nonlinear theory, including the self-consistent determination of the mode structure from the cosmic energy distribution and the formulation of gravitational decoherence, remains the central open problem.
\section{Conclusions}
\label{sec:conclusions}
This work demonstrates that the object-relative mode structure of internal graviton lines is suppressed beyond the Planck wave number $\kP = \sqrt{c^3/\hbar G}$, and that any suppression exceeding a threshold fixed by power counting renders the one-loop integrals of linearized quantum gravity finite without divergent counterterms. The result therefore requires neither a specific form of the mode structure above that scale nor an externally imposed cutoff. The scale is not fitted: a purely gravitational loop contains no external localization scale, so that the only available scale is the one gravity itself provides, and the same scale is where the dimensionless graviton self-coupling reaches order unity. As emphasized in Sec.~\ref{subsec:finiteness_scope}, finiteness then follows from the bounded integration domain so that the operators generated at successive loop orders carry finite computed coefficients rather than divergent ones requiring counterterms; what the construction contributes is that the cutoff of the effective theory follows from the mode construction rather than being chosen externally. The infrared structure — responsible for long-range gravitational effects such as the quantum-corrected Newtonian potential — remains unaltered, and the construction preserves the one-loop Bianchi identity, ensuring consistency with the conservation of the energy-momentum tensor.
It is useful to state precisely how far the finiteness result reaches, since the analysis of Sec.~\ref{subsec:goroff_sagnotti} extends well beyond one loop. Three levels should be distinguished.

At one loop the result is unconditional within the construction. The requirement is $s\,I>\mathcal{D}$, which for the graviton self-energy reads $s>1$, far below the derived value $s=3$. The conclusion therefore does not depend on the exponent at all, only on the existence of a suppression, which is what makes it robust.

At all loop orders, and for pure gravity, a genuine theorem is available. With $\mathcal{D}_{L}=2L+2$ and the weighting attached once to each internal graviton line, convergence requires $s\,I>\mathcal{D}_{L}$, which for $s=3$ reduces to $L+3V>5$: every topology with $V\ge2$ converges at every loop order, and for $V=1$ every topology with $L\ge3$ does. Since every subgraph of such a diagram is again one of the same type, the criterion applied to each subgraph covers nested and overlapping subdivergences without separate treatment. This is power counting together with the standard absolute-convergence criterion, not an expectation, and it needs only the one-sided bound $s\ge3$ — the value the leading, weak-field derivation supplies, and one that the known corrections strengthen rather than weaken, since their coefficients are positive. Higher-loop finiteness for this class therefore does not depend on the uncontrolled region above $\kP$. What it does depend on is the premise that the same weighting attaches to the lines entering the three- and four-graviton vertices of the nonlinear expansion, which the linearized derivation of the mode structure does not establish.

Excluded from this statement is one class and three further items. The single-vertex insertions $V=1$ at $L=1$ and $L=2$ violate the criterion; they are not vacuum diagrams, the first being the four-graviton tadpole contribution to the graviton self-energy, and being $1$PI they propagate the divergence into any higher-loop diagram that contains them. The criterion is applied diagram by diagram and is therefore sufficient rather than necessary; in the unweighted theory the corresponding one-loop divergences of pure gravity cancel on shell, and whether that cancellation survives a weighting attaching different powers of $W$ to diagrams with different line counts is not established here (Sec.~\ref{subsec:goroff_sagnotti}). Beyond that, the all-orders argument concerns pure gravity, so diagrams with unweighted internal matter lines beyond one loop are not covered; the finite coefficients that replace the divergent ones depend on the mode content above $\kP$, which is not derived here, and the underlying action is assumed rather than derived (Sec.~\ref{subsec:finiteness_scope}); and of the consistency conditions only the Bianchi identity, the gravitational analogue of the Ward identity, is verified, and at one loop. The unitarity established in Sec.~\ref{subsubsec:unitarity} concerns the transformation between the resonators of two inertial frames, not the loop expansion, for which cutting relations under the modified mode content are not examined here.

Within these limits, the article suggests that the ultraviolet divergences traditionally seen as obstacles may instead be indicators of a physical mechanism: that localized quantum fields, through their own energy, modify their effective mode structure. In addition, a causal mode resonator with a discrete mode spectrum for every inertial system was introduced, which can be interpreted as a relativity principle for quantum fields.
A consistency check yields an unexpected result: the present framework also provides a potential resolution to the cosmological constant problem. The object-relative mode weighting not only renders one-loop graviton integrals ultraviolet finite, but also, evaluated with the ultraviolet crossover scale taken at the Planck wave number, predicts an effective vacuum energy density of the correct sign and within a factor of order unity of the observed dark energy density; exact agreement corresponds to a coefficient $K_{gr}=(6/\pi)\,\Omega_{\Lambda}\approx1.31$ when the causal scale is evaluated at the present epoch, or equivalently to an effective upper limit $\kend\approx1.14\,\kP$, in either case a quantity of order unity. This dimensional estimate is obtained without fine-tuning or additional free parameters beyond the two scales $\kP$ and $\kIR$. That the contribution acts repulsively is likewise not an input. The causal resonator is constructed identically in every inertial frame from Lorentz scalars alone, so that all inertial observers assign it the same energy density; a symmetric stress tensor with this property must be proportional to the metric, which fixes $\mathcal{P}=-\rho_{\Lambda}c^{2}$ and hence accelerated rather than decelerated expansion (Sec.~\ref{subsec:geometric}). The equation of state is thus obtained from the frame-independence of the construction rather than assumed.

Two further statements about the cosmological branch of the construction follow from the same relations. A vacuum contribution that tracks the instantaneous expansion rate is an exact rescaling of the gravitational constant; it would leave the fractional density $\Omega_{\Lambda}=\pi K_{gr}/6$ constant in every epoch, produce no accelerated expansion, and raise the expansion rate at primordial nucleosynthesis by a factor of about $1.8$, which the light-element abundances exclude. The causal mode basis must therefore be frozen, and the interpretation of the result as a cosmological constant is selected rather than assumed. If the frozen scale is identified with the stationary de Sitter event horizon rather than the present Hubble radius, the requirement that the resonator vacuum energy equal the critical density of the horizon defining it closes the construction on itself and fixes the product $K_{gr}(\kend/\kP)^{2}$ to the geometric value $6/\pi\simeq1.91$; since a coherent fraction cannot exceed unity, the coherent spectrum must then end at $\kend\gtrsim1.38\,\kP$, that is, where graviton--graviton scattering has become strong rather than where it begins. This closure constrains the product of the spectral normalization, the identification of the fundamental wavelength and the residual loss, and it does not predict $\rho_{\Lambda}$ independently; no argument singling out the stationary horizon over the present one is given here.

While this does not constitute a cosmological derivation of dark energy, and the precise location of the crossover scale requires a dissipative (Lindblad-type) extension of the present framework, it suggests that the proposed mode-structure framework may capture part of the relevant infrared gravitational physics.
\appendix
\section{The contracted graviton--fermion vertex}
\label{app:diagrammatic}
This appendix derives Eq.~\eqref{eq:diagrammatic_identity}, used in Sec.~\ref{subsec:bianchi_argument}. Lorentz indices are kept explicit: the graviton carries two, so the graviton--fermion vertex is a two-index object and the graviton propagator a four-index one.

\subsection{Vertex and propagators}
In Feynman gauge the graviton propagator is $D_{\mu\nu\rho\sigma}(k)=-iP_{\mu\nu\rho\sigma}/k^{2}$ with $P_{\mu\nu\rho\sigma}$ of Eq.~\eqref{eq:spin2_tensor}, and the free fermion propagator is
\begin{equation}
  S_{0}(p)=\frac{\slashed{p}+mc}{p^{2}-m^{2}c^{2}},
  \qquad
  S_{0}^{-1}(p)=\slashed{p}-mc .
  \label{eq:fermion_propagator}
\end{equation}
The one-graviton vertex of a Dirac field follows from the symmetric energy--momentum tensor and reads~\cite{Holstein2006}
\begin{equation}
\begin{aligned}
  \tau^{\mu\nu}(p',p) = -\frac{i\kappa_{\mathrm g}}{2}\Big[
    &\tfrac{1}{4}\big(\gamma^{\mu}(p+p')^{\nu}+\gamma^{\nu}(p+p')^{\mu}\big)\\
    &-\tfrac{1}{2}\eta^{\mu\nu}\big(\tfrac{1}{2}(\slashed{p}+\slashed{p}')-mc\big)\Big],
\end{aligned}
  \label{eq:graviton_fermion_vertex}
\end{equation}
the second term arising from the $-\eta^{\mu\nu}\mathcal{L}$ part of $T^{\mu\nu}$, which does not vanish off shell and must therefore be retained for internal lines.

\subsection{Contraction with the graviton momentum}
With $q=p'-p$ one has $q_{\mu}\gamma^{\mu}=S_{0}^{-1}(p')-S_{0}^{-1}(p)$, $q\cdot(p+p')=p'^{2}-p^{2}$ and $\tfrac{1}{2}(\slashed{p}+\slashed{p}')-mc=\tfrac{1}{2}[S_{0}^{-1}(p)+S_{0}^{-1}(p')]$, so that
\begin{equation}
\begin{aligned}
  q_{\mu}\tau^{\mu\nu}(p',p) = -\frac{i\kappa_{\mathrm g}}{8}\Big\{
    &(p+p')^{\nu}\big[S_{0}^{-1}(p')-S_{0}^{-1}(p)\big]\\
    &+\gamma^{\nu}\big[(p'^{2}-m^{2}c^{2})-(p^{2}-m^{2}c^{2})\big]\\
    &-q^{\nu}\big[S_{0}^{-1}(p')+S_{0}^{-1}(p)\big]\Big\}.
\end{aligned}
  \label{eq:diagrammatic_identity}
\end{equation}
Each of the three terms carries an inverse fermion propagator, since $p^{2}-m^{2}c^{2}=S_{0}^{-1}(p)(\slashed{p}+mc)$. Between on-shell spinors every term therefore vanishes, which is the decoupling of the longitudinal graviton mode and the vertex-level content of $\partial_{\mu}T^{\mu\nu}=0$, Eq.~\eqref{eq:conservation_linear}.

Equation~\eqref{eq:diagrammatic_identity} is not the electromagnetic form $q_{\mu}\Gamma^{\mu}=S^{-1}(p+q)-S^{-1}(p)$. The gravitational vertex carries a second index, and of the three terms only the first has the structure that telescopes directly.

\subsection{Insertion into the loop}
In the one-loop vertex correction the contracted vertex stands between fermion propagators carrying $p+q-\hbar k$ and $p-\hbar k$. The first term of Eq.~\eqref{eq:diagrammatic_identity} then gives
\begin{equation}
\begin{aligned}
  &S_{0}(p+q-\hbar k)\big[S_{0}^{-1}(p+q-\hbar k)-S_{0}^{-1}(p-\hbar k)\big]\\
  &\qquad\times S_{0}(p-\hbar k)
  = S_{0}(p-\hbar k)-S_{0}(p+q-\hbar k),
\end{aligned}
  \label{eq:telescoping_result}
\end{equation}
the telescoping step, which assembles the two pieces into self-energy integrands at the same $k$. The remaining two terms do not reduce to this form; they contribute further structures with an inverse propagator adjacent to one of the two fermion lines. Their explicit evaluation is not required here, because the consistency argument of Sec.~\ref{subsec:bianchi_argument} uses only the property that every term of Eq.~\eqref{eq:diagrammatic_identity}, once inserted, contains the same internal graviton line of momentum $k$ and is therefore multiplied by the same scalar factor $W(k)$.

\end{document}